\documentclass{aa}

\usepackage{graphicx}
\usepackage{natbib}
\usepackage[colorlinks,urlcolor=magenta,citecolor=blue,linktocpage,breaklinks]{hyperref}
\usepackage[varg]{txfonts} 
\usepackage{threeparttable}

\newcommand{\dftwoPA}{$51.1$}
\newcommand{\dftwoAR}{$0.83$}

\newcommand{\dffourPA}{$36$}
\newcommand{\dffourAR}{$ 0.83$}

\begin{document} 

\title{Ultra-deep imaging of NGC1052-DF2 and NGC1052-DF4 to unravel their origins}

\author{Giulia Golini, 
      \inst{1,2}
      Mireia Montes\inst{1,2}, Eleazar R. Carrasco\inst{3}, Javier Román\inst{4,1,2}, Ignacio Trujillo\inst{1,2}}

\institute{Instituto de Astrofísica de Canarias,c/ Vía Láctea s/n, E38205 - La Laguna, Tenerife, Spain 
\and Departamento de Astrofísica, Universidad de La Laguna, E-38205 - La Laguna, Tenerife, Spain 
\and Gemini Observatory/NSF’s NOIRLab, Casilla 603, La Serena, Chile
\and Kapteyn Astronomical Institute, University of Groningen, Landleven 12, 9747 AD Groningen, The Netherlands
     }

\date{Received 9 November 2023 / Accepted 25 January 2024 }

\abstract
{A number of scenarios have been proposed to explain the low velocity dispersion (and hence possible absence of dark matter) of the low surface brightness galaxies NGC1052-DF2 and NGC1052-DF4. Most of the proposed mechanisms are based on the removal of dark matter via the interaction of these galaxies with other objects. A common feature of these processes is the prediction of very faint tidal tails, which should be revealed by deep imaging ($\mu_g>$ 30 mag/arcsec$^2$). Using ultra-deep images obtained with the Gemini telescopes, about 1 mag deeper than previously published data, we analyzed the possible presence of tidal tails in both galaxies. We confirm the presence of tidal tails in NGC1052-DF4, but see no evidence for tidal effects in NGC1052-DF2, down to surface brightnesses of $\mu_g$=30.9 mag/arcsec$^2$.  We therefore conclude that while the absence of dark matter in NGC1052-DF4 could be attributed to the removal of dark matter by gravitational interactions, in the case of NGC1052-DF2 this explanation seems less plausible, and therefore other possibilities such as an incorrect distance measurement or that the system may be rotating could alleviate the dark matter problem.}

\titlerunning{Ultra-deep imaging of DF2 and DF4}
\authorrunning{Golini et al.}
\maketitle

\section{Introduction}

NGC1052-DF2\footnote{This galaxy was previously named KKSG04, PGC3097693, and [KKS2000]04 \citep[see, e.g.,][]{2000A&AS..145..415K}. } and NGC1052-DF4 \citep{discoverydf2, discoverydf4}, two low surface brightness galaxies ($<$$\mu_{V}>_{eff}$ $>$ 25 mag/ arcsec$^2$) in the line of sight of the NGC1052 group\footnote{The interested reader can find the location of these two galaxies in relation to NGC1052 and NGC1042 in Fig. 1 of \citet{muller2019}.}, have attracted the attention of the extragalactic community in recent years.
Both galaxies have large effective radii (R$_{eff}$$>$1.5 kpc) for their typical stellar masses ($\sim$10$^8$ $M_{\odot}$; \citealt{discoverydf2,discoverydf4}).
Interestingly, the very low velocity dispersion of their globular cluster (GC) systems (4-10 km/s) suggests that the amount of dark matter they contain is either very small or negligible. The possibility of a low dark matter content sparked a lively debate, much of which focused on how best to estimate the intrinsic velocity dispersion of NGC1052-DF2 \citep[see, e.g.,][]{2018Martin,vandokkum2018b,2019fensch,Emsellem2019}, the distance to these objects \citep[see, e.g.,][]{2019Trujillo,monellitrujillo,zonoozi}, and/or the possibility that these galaxies are rotating \citep[see, e.g.,][]{lewis2020}.
While the above concerns can mitigate the problem of the absence of dark matter, especially in the case of NGC1052-DF4, where the velocity dispersion of the GC system is extremely small, the absence of dark matter remains a strong hypothesis.

Numerous hypotheses have been proposed to explain the peculiar properties of these galaxies, but none has provided a definitive answer \citep[see, e.g.,][]{2018Bennet,2018Famaey,2018Kroupamond,2019fensch, 2022vandokkum}. 
Among the proposed formation mechanisms, the most compelling explanation is that these galaxies had a typical dark matter content, which they subsequently lost through interactions with neighboring objects \citep{ogiya2018, 2020Jackson, maccio2021, ogiya2022}. 

Since the distribution of dark matter is more extended than the stars, during a tidal interaction with a more massive neighbor, the galaxy first loses its dark matter and then its stars begin to be stripped along two symmetrical tidal tails emanating from two opposite sides of the galaxy. These tails exhibit an S-shaped structure with a width equal to the diameter of the disrupted object \citep[e.g.,][]{Johnston2002,2022Moreno}.
Recent simulations \cite[see, e.g., ][]{2022Moreno,2023KATAYAMA} show that if the dark matter in NGC1052-DF2 and NGC1052-DF4 has been stripped by gravitational interaction with a more massive object, then the tidal tails should be visible in ultra-deep imaging, reaching a surface brightness of $\mu_V$$\sim$30 mag/arcsec$^2$ (even if the interaction took place several gigayears ago).

Therefore, the key to testing these tidal stripping scenarios is to have very deep imaging of the galaxies under study. Several papers in the past have examined these two galaxies with deep data to analyze their outskirts and surroundings, looking for (among other things) tidal features \citep[see, e.g.,][]{muller2019,2019Trujillo, 2021roman}. 
The tidal stripping scenario for NGC1052-DF4 has recently been confirmed observationally by \citet{Montes2020} using the IAC80 telescope and by \citet{Keim2022} with deep imaging from the Dragonfly Telephoto Array. Both studies have independently shown, using very deep imaging, that NGC1052-DF4 has the characteristic S-shaped structure indicative of tidal interaction. 
While the two teams agree on the tidal distortion of the galaxy, the massive galaxy causing the distortion is a matter of debate.  \citet{Montes2020} argue that the observed tidal tails point to NGC1035, a medium-sized spiral very close to NGC1052-DF4 in projection, which also shows signatures of being distorted. On the other hand, \citet{Keim2022} suggest that the massive galaxy producing the tidal features of NGC1052-DF4 is NGC1052.

For the other low-mass galaxy, NGC1052-DF2, the situation is far from clear. \citet{Montes2021}, using deep images taken with the \textit{Isaac Newton} Telescope (INT), find no evidence for tidal tails in the outer parts of the galaxy. Instead, they find a surface brightness profile that is cored in the inner region and decreases exponentially in the outer part. 
They argue this could indicate the presence of a low-inclination disk, supporting the evidence for rotation suggested by \citet{lewis2020}.  However, \citet{Keim2022} interpret this core-like shape in the innermost region of the galaxy as a signature of the tidal interaction of NGC1052-DF2 with NGC1052. 

While the images obtained so far for the two galaxies are very deep ($\mu_V$$\sim$29-29.5 mag/arcsec$^2$; 3$\sigma$ in 10\arcsec$\times$10\arcsec\ boxes), they seem insufficient to determine the presence of tidal disruption features. The aim of the present work is to probe the galaxies with images obtained by the Gemini telescopes that are about 1 mag deeper than previous data published. Following the motivation of \citet{2022Moreno} and \citet{2023KATAYAMA}, we hope that with these ultra-deep data the presence of tidal distortions, if present, can be unambiguously revealed.

The paper is organized as follows.
Section \ref{sec:data} provides an overview of the observational strategy employed and outlines the data reduction pipeline used to obtain our ultra-deep images.
Section \ref{sec:analysis} presents the results of our analysis, including the surface brightness profiles of NGC1052-DF2 and NGC1052-DF4. In Sect. \ref{sec:discussion} we compare our results with previous research. Finally, we summarize our conclusions in Sect. \ref{sec:conclusions}. All magnitudes in this paper are given in the AB magnitude system. 

\addtolength{\textheight}{\topskip}

\section{Data}
\label{sec:data}

The ultra-deep images of NGC1052-DF4 and NGC1052-DF2 were obtained with the two Gemini Multi-Object Spectrographs (GMOS)\footnote{\url{http://www.Gemini.edu/instrumentation/gmos}} mounted on the Gemini North and South telescopes \citep{2004hook}. 
The GMOS cameras cover a $5^\prime\times 5^\prime$ square area with truncated corners. Each camera consists of three adjacent charge-coupled device (CCD) sensors, and each of the three CCDs is read by four amplifiers. The first two amplifiers of CCD1 and the last two of CCD3 are outside the imaging field of view (FoV) and, therefore, they do not contribute to the final dataset. 
On the new Hamamatsu CCDs for GMOS-S, the gaps between the CCDs are 4.88\arcsec\ wide (61 pixels in unbinned mode), while the gaps between the CCDs for the GMOS-N Hamamatsu CCDs are 5.4\arcsec\ wide (67 pixels in unbinned mode).  We worked in 4 $\times$ 4 pixels binned mode resulting in a final pixel scale of  0.32\arcsec/pix.
Due to the specific configuration of the telescope, vignetting from the On-Instrument WaveFront Sensor (OIWFS), which provides guidance and tip-tilt correction information to the telescope control system, appears as an artifact on the data, which is properly accounted for as described in the following Sect. \ref{sub:observationalstrategy}.

NGC1052-DF2 was observed on October 28, 2021, with GMOS-S (r filter) and on November 8, 2021, with GMOS-N (g filter).
The data of NGC1052-DF4 (r filter) were obtained on October 31, 2021, with GMOS-S. The mean seeing during the observations was 1.3\arcsec\ in r filter for DF4 and 0.88\arcsec\ and 1.2\arcsec\ in r and g filters, respectively, for DF2.

To perform photometric calibration of Gemini data and check the consistency of our results, we used public Sloan Digital Sky Survey (SDSS) and Dark Energy Camera Legacy Survey (DECaLS) data that are currently available for the objects.
We retrieved the SDSS g- and r-band imaging data from the DR14 SDSS \citep{2018sdss} Sky Server. The magnitude zero point for the SDSS dataset is 22.5 mag and the pixel scale is 0.396$\arcsec$/pixel.
We downloaded g- and r-band optical data from DECaLS (Proposal ID 2014B-0404; PIs: David Schlegel and Arjun Dey) archive with a zero point of 22.5 mag and a pixel scale of 0.262$\arcsec$/pixel. We used the DECaLS DR10 data from bricks 0405m085 and 0405m083 for NGC1052-DF2 and NGC1052-DF4 fields, respectively.

\subsection{Observational strategy}
\label{sub:observationalstrategy}
\begin{table*}
    \centering
    \caption{Characteristics of the Gemini observing run conducted to achieve ultra-deep imaging of NGC1052-DF2 and NGC1052-DF4.  The table shows the exposure time and the number of frames taken during each night for both the sky and the on-source images. The surface brightness limit of the data (3$\sigma$; $10\arcsec \times 10\arcsec$ boxes) and the camera used are also given.
}
    \begin{tabular}{c|c|c|c|c|c|c|c|c}
         \hline\hline
         Object & Camera & filter & $t_{exp,obj}$ & $n_{frames,obj}$ & $t_{exp,sky}$ & $n_{frames,sky}$ & $t_{total,obj}$ & $\mu_{limit}$ (3$\sigma$; $10\arcsec \times 10\arcsec$ boxes) \\
         \hline
         NGC1052-DF4 & GMOS-S & r & 150s & 36 & 50s & 25 & 1h 30min & 30.6 mag/arcsec$^2$ \\
         NGC1052-DF2 & GMOS-S & r & 150s & 32 & 50s & 25 & 1h 20min & 30.5 mag/arcsec$^2$ \\
         NGC1052-DF2 & GMOS-N & g & 300s & 18 & 60s & 25 & 1h 30min & 30.9 mag/arcsec$^2$ \\
         \hline
    \end{tabular}
    \label{tab:exposures}
\end{table*}

To reach the depths we are aiming for in this work (i.e., limiting surface brightness of around $\mu_g$ $\sim$ 31.0 mag/arcsec$^2$, 3$\sigma$ in $10\arcsec \times 10\arcsec$), we need to deal with several observational biases that can affect deep imaging, such as scattered light, saturation, or ghosts. In addition, we need an accurate estimation of the flat field and a careful treatment of the sky background. To this end, we designed a special observing strategy for the Gemini telescopes that consists of a dithering pattern with offsets of 10\arcsec. The offset is needed to fill the gap between the CCDs in the cameras and to minimize scattered light from the telescope optics. This offset also ensures that we minimize the errors introduced by quantum efficiency and differences between the amplifiers, since the galaxy always overlaps on the central CCD (CCD2). The final dithering pattern around both galaxies is shown in Figs. \ref{fig:dithering-df2} and \ref{fig:dithering-df4}. NGC1052-DF2 is not centered in the Gemini FoV to better characterize the scattered light from the bright star to the east (see Sect. \ref{subsub:scatterlight}).

The differences between the night sky and the twilight illumination can lead to subtle differences in the flat field. The twilight light gradient can introduce unwanted gradients in our data. Therefore, the twilight flats provided by the observatory are unsuitable for our purpose. A more accurate flat-field correction is obtained using the same scientific exposures obtained with a large displacement dithering pattern \citep[typically of the size of the object under observation; see ][]{trujillofliri2016}. 

Unfortunately, the main limitation of using this approach with the Gemini telescopes revolves around the size of the OIWFS patrol field.
When making large offsets (>10\arcsec), the risk is to push the guide star beyond the patrol field's boundaries. Consequently, employing the same science field with extensive offsets could lead to loss of guiding, potential vignetting in one or more images, and a reduction in the effective FoV, resulting in a smaller area with a higher signal when stacking images.
This is true for NGC1052-DF4, where the Guide star is located in the external border of the OIWFS patrol field while in the case of NGC1052-DF2, the Guide star was closer to the GMOS FoV.

To mitigate this problem, every six science observations of the galaxy, the telescope is pointed at a nearby empty region of the sky ($\sim$7$\arcmin$ away from the galaxy). In such a region, we took five unguided exposures on sky. With this sky data we created the flat field.  Using this technique, we retrieved $\sim$20 min of sky observations for each night to build the flat field from the data.

At the end of the observing run, we got 1h30min for NGC1052-DF4 (r filter), 1h20min for NGC1052-DF2 (r filter), and 1h30min for NGC1052-DF2 (g filter).
Table \ref{tab:exposures} lists the exposure times for each filter and the number of frames, together with the limiting surface brightness depths of the data. Bias frames were taken at the end of the nights as part of the observing programs. They consist of 20 bias frames per CCD.

\subsection{Frame processing}
\label{sub-general-processing}

In this subsection, we explain in detail how the sky images and the science images are processed and assembled to obtain the final co-addition.
The main steps of the process are:
(i) CCD assemblage (gain correction), (ii) the derivation of the flat field using sky images, (iii) custom processing of galaxy images, and (iv) mosaic co-addition.

The pipeline makes effective use of the \texttt{Gnuastro} Software \citep{gnuastro} and its tasks.

\subsubsection{Bias and gain correction}
\label{sub-bias-gain}

First, for each of the three CCDs and for each individual image (both on-source and on-sky), we generated a bad pixel mask. This mask account for various cosmetic defects present on the CCDs, such as the readout failure of the detector, over-scan regions, and bad columns.
In practice this mask implies changing these pixels to not-a-number values.

Also, we masked all pixels with values greater than $5\times 10^4~\rm{ADUs}$ to account for the sensor nonlinearity.
After this, we produced bias-corrected images.
To do so, we created a master bias (for each CCD) by combining the 20 bias frames we have for each of the cameras and nights with a resistant mean (using a 3$\sigma$ rejection).

After subtracting the master bias, we flattened the response of the CCDs, since they are an array of four different amplifiers, each with a different gain. For this reason, we used the following strategy to obtain the most homogeneous image possible. We started by selecting one of the central amplifiers as a reference for the other amplifiers.
We masked all the sources of the images to avoid their light contamination. Also, we assumed continuity of the background between two adjacent amplifiers. This is a reasonable assumption, since we do not expect strong changes in the illumination of the camera, nor background relevant gradients between adjacent pixels on the sky. In our case, the background pixels of the reference amplifier are selected from a column at the edge with a width of 10 pixels.  The median of these pixels is our reference background for that amplifier. We did the same estimation for the adjacent amplifier using the adjacent column. We then calculated the ratio between the two quantities to multiply the gain we want to use with respect to the reference gain. Once these two amplifiers are corrected, the same procedure is applied to the other amplifiers of the CCD. 
 While the above procedure nicely flatten the gain levels at the boundaries of the amplifiers for each individual image, it still leave some doubt about how good our initial hypothesis is about the continuity of camera illumination and background light between adjacent amplifiers. For this reason, we tried to minimize any potential deviation from this hypothesis from frame to frame by applying a final gain correction that has the information of all individual frames of a given night. We assumed that the gain of the amplifier remains constant during the few hours of observation of the source and the sky. In this case, the correction that we finally applied to leverage out the amplifiers is the median of all the corrections between any two amplifiers that we find from one frame to another. We find that the ratio between the backgrounds applied to do the flattening correction is extremely stable over the observation run, with a variability around 0.1\%.

\subsubsection{Flat-field correction}
\label{sub:flat-field}

In this section, we explain how we combined our on-sky images to build the flat field in a two-step process, following a strategy similar to that described in \cite{LIGHTSs}.
Once the raw sky images are bias- and gain-corrected (as described in Sect. \ref{sub-bias-gain}), we computed a crude flat-field image using a median of the normalized sky images for each CCD, without any masking. In order to follow the illumination pattern of the entire GMOS camera, the normalization of each individual CCD that makes up the camera does not use the pixels of the entire CCD, but only the portion of the CCD that is expected to be illuminated with the same efficiency. In practice, this means that each CCD image is normalized using the area inside a circular annulus of 100 pixels width centered on the central pixel of CCD2 (i.e., the central CCD of the camera) and radius of 420 pixels. With this radius, we make sure that the annulus covers the entire FoV of the camera (CCD1 + CCD2 + CCD3). Then, for each CCD, we calculated the resistant median of the pixels within the corresponding annulus section.  This value was used to normalize the flux in each frame before they were added together to create the flat field.

The rough flat fields are the combination using the sigma-clipped (3$\sigma$) median of the normalized sky images. Then, to get an improved master flat field, we used the corrected (by the first rough flat field) sky CCDs images to build a mask of the astronomical sources using \texttt{NoiseChisel} \citep{noisechisel_segment_2019}. This was done because even with a first rough flat-field correction the images show the astronomical sources very well. With these pixels masked, we next built a much more accurate flat field, following the previous steps again.

Having the master flat field generated, each bias- and gain- corrected science frame was divided by the master flat field. These master flat-field-corrected images were then used to generate better masks of the astronomical sources of the science images. These masks were therefore used to improve the previous step related to the gain correction. We, consequently, repeated the previous steps to improve our reduction. With the very final version of the master flat field, we performed the flat-field correction of our science frames. 

Finally, to avoid vignetting problems toward the corner of the detectors, we removed from the images all the pixels where the master flat field have an illumination lower than 90\% of the central pixels.


\subsubsection{Astrometry} 
\label{subsub:astrometry}

In the following, we use the entire assembled frame of the camera (i.e., CCD1 + gap + CCD2 + gap + CCD3) as the unit of work. To construct the complete pointing of the camera, we took the header of the preprocessed images with THELI \citep{2013THELI}, a data processing pipeline for astronomical images, to have a first astrometric solution. We then used the \textit{makenew} operator in \texttt{Gnuastro} to create the gap image between the CCDs (with different widths depending on the camera) and we stitched the CCDs and gaps together in a single fits file with a single extension.  On the images we ran SExtractor \citep[v.2.25.2; ][]{1996A&AS..117..393B} to generate the input catalogs for SCAMP \citep[v.2.10.0; ][]{2006ASPC..351..112B} to improve the astrometry and the distortion of the camera. SCAMP was run twice, once with -MOSAIC TYPE \footnote{Preprocessing technique applied to multi-extension FITS catalogs of focal plane mosaics.} set to LOOSE and once in UNCHANGED mode. In the initial iteration, the CCD images are considered as separate and mechanically unconnected devices. Pattern matching is conducted independently for each extension. In the subsequent iteration, no preprocessing is assumed and the new distortion headers are computed on the basis of the first iteration. This approach allows for a more accurate characterization of distortions, achieving a precision of 0.01\arcsec. These new headers are our final world coordinate system (WCS) solutions for each pointing.

\subsubsection{Sky subtraction}
\label{subsub:sky}

Sky subtraction is a crucial step in reducing low surface brightness data; if not done correctly, it can introduce artificial gradients and affect the faint objects under study. In this work, since the FoV of the camera is not very large, we worked under the hypothesis that the sky in each frame can be well described by a single constant. 

To compute the sky, we first masked the brighter sources and the diffuse light with \texttt{Noisechisel}.
To account for potential non-detections by \texttt{Noisechisel} of diffuse light surrounding bright objects, a manual mask is applied after the initial \texttt{Noisechisel} detection process.

In the case of the NGC1052-DF2 field, we added a manual circular mask around the bright star (i.e., HD16873 with m$_V$=8.34 mag and G8III, located at RA=40.5395 deg and Dec=-8.3768 deg; see Fig. \ref{fig:dithering-df2}) close to our FoV. 
The manual mask has a radius of 4$\arcmin$. This radius was chosen so that at this radial distance the expected contribution of scattered light from this bright star is $\sim$28.5 mag/arcsec$^2$. This is roughly the limiting surface brightness depth of each of our individual exposures. Consequently, we used such masks to try to minimize the effect of the scattered light on our determination of the sky background.
For the NGC1052-DF4 field instead, we built an extra elliptical mask around the galaxy NGC 1035 with a position angle (PA) of 20 degrees (clockwise from the north axis), an axis ratio of 0.28, and a maximum major axis of 4$\arcmin$.  In the same field, we also added a circular mask with a radius of 1$\arcmin$ over the nearest bright star (RA=39.7778; Dec=-8.1121) toward the west side of the galaxy. In both fields, we computed the sky values of each frame as the 3$\sigma$ clipped median of the non-masked pixels and subtract these values from the data. 
Before removing the sky value, we visually inspected \texttt{Noisechisel}'s output on the distribution of background pixels within the image to ensure that our assumption of a constant value is accurate. We performed a visual inspection to identify any remaining gradients in the image. No gradients were observed.

\subsubsection{Photometry}
\label{subsub:photometry}

Once the sky value has been removed from each individual exposure, we converted the Gemini pixel values from analog digital units (ADUs) into into linear physical units of nanomaggies\footnote{\url{https://www.sdss3.org/dr8/algorithms/magnitudes.php}}.
To do this, we used the SDSS data covering the same FoV around our target.
For both the Gemini images and the SDSS mosaic, we stored a catalog of the flux of the sources in the field using circular apertures with a diameter of 2\arcsec\ with \texttt{Gnuastro} \textit{astmkcatalog}. These catalogs were then checked against the \textit{Gaia} DR3 star catalog to make sure that we were using verified point-like sources for the photometry.  We stored only the parameters of the stars that are not saturated in the Gemini data (i.e., we used only stars with magnitudes g$>$18 and r$>$17.8 mag) in order to minimize errors in the estimation of the flux of the sources. In addition, we only used stars with high enough signal-to-noise in SDSS images, that is, g and r magnitudes of less than 21 mag. For all unsaturated stars, we computed the ratio of the flux in the SDSS circular aperture ($F_{SDSS, ap}$) to the flux in the Gemini aperture ($F_{Gem, ap}$) for each science exposure. Then we multiplied the Gemini image ($im_{G}$) by the resistant median of all these ratios. At the end of the process we have the pixel values of each pointing in units of nanomaggies ($im_{G,n}$).

\subsubsection{Image co-addition}
\label{subsub:weightedmean}

In the final step of data reduction, the individual exposures are added together to form the final mosaic. The following steps are motivated by the fact that the conditions of our observations change during the night. For example, the airmass increases or decreases depending on the trajectory of the target in the sky as it moves further or closer to zenith (see Fig. \ref{ima:airmass}). In addition, other meteorological phenomena such as the passage of a cloud can change the brightness of the sky value. These local variations manifest themselves as a degradation of sky quality: the higher the standard deviation of the (photometrically calibrated) sky background pixel values, the worse the image quality. To account for these effects and improve the accuracy of the low surface brightness features, the images taken under better conditions should weigh more than the images taken under worse conditions. 

To carry out the co-addition of the individual frames, we used a weighted average to stack the images, where the weight assigned to the image is the ratio of the standard deviation of the sky of the best exposure to the standard deviation of the sky of the i$^{th}$ frame. However, before stacking the data in this way, it is necessary to mask those pixels that have been affected by unwanted signals such as cosmic rays. These pixels are identified using the following procedure. First we used \texttt{Swarp} \citep{swarp2010} to put the images into the same astrometric solution. Then we created a mask to reject the pixels that have a value that deviates from the resistant mean (3$\sigma$ rejection is sufficient for our purpose) of all pixels at the same WCS location, in all frames. We then masked the outliers and stacked the data using this \texttt{sigclip}-weighted mean.

The co-added image is significantly deeper than any single image, and therefore a large number of very low surface brightness features, previously invisible, emerge from the noise. These features subtly affect the sky determination of our individual science images, and for this reason it is necessary to mask these regions and repeat the entire sky determination process on the individual exposures with an improved mask. In short, we repeated the sky estimation (and subsequent reduction steps) described above on the individual images using the improved masks generated by this first data co-addition.

\subsubsection{Scattered light removal in the NGC1052-DF2 field}
\label{subsub:scatterlight}

\begin{figure*}
    \centering
    \includegraphics[width=\textwidth]{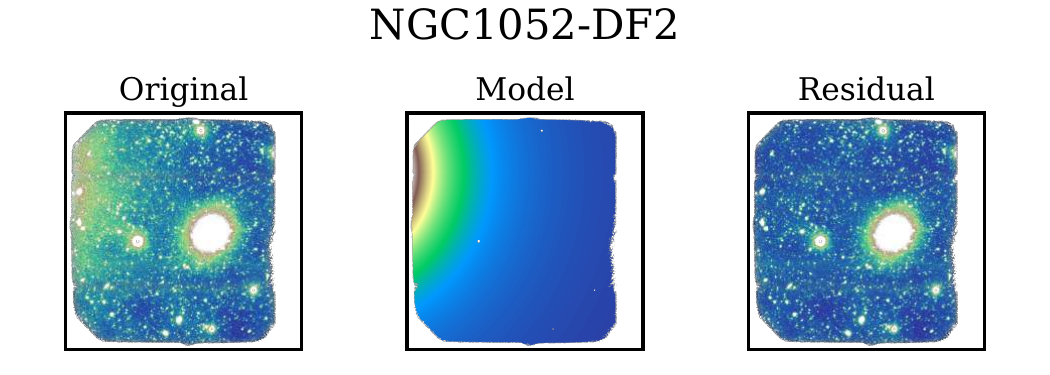}
    \caption{Three stages of subtraction of the scattered light emitted by the star HD16873 near NGC1052-DF2. The panels depict the complete mosaic of the field (5\arcmin$\times$5\arcmin) using g-filter Gemini data. The left panel illustrates the original stacked image, while the middle panel shows the model fitted to the starlight. Subtracting the star model from the original image yields the residual image shown in the right panel.}
    \label{ima:fitg}
\end{figure*}

As mentioned above, one of the main difficulties in obtaining a high quality image for low surface brightness science in the field around NGC1052-DF2 is the presence of a very bright star (HD16873) close to the galaxy. Therefore, we needed to develop a specific strategy to remove the light gradient produced by the scattered light of such a bright source outside the FoV of our co-added image. Figure \ref{ima:fitg} illustrates the three stages of subtraction of the scattered light emitted by the star HD16873 near NGC1052-DF2. The left panel of the Fig. \ref{ima:fitg} shows that the light from the star HD16873 manifests itself as a clear excess of light on the northeast side of the final reduction of NGC1052-DF2. 

To remove the light pollution from the bright star, we did the following. First, we masked all major bright sources in the final Gemini mosaic with \texttt{Noisechisel}. Second, only by using the non masked pixels in the Gemini stacked image, we obtained the radial profile of the scattered light produced by that star using circular apertures centered on the star position (i.e., RA = 40.5395 deg, Dec = -8.3768 deg).
In practice, we get the light profile of the star only in the Gemini FoV region (from $\sim$2.4\arcmin\ to $\sim$9\arcmin).
Such a profile is fitted with a power law shape $\beta$r$^{\alpha}$. We find that the slope of the light due to the scattered light from the star is well described by a power law with $\alpha$ = -2.6 for the g filter and $\alpha$ = -2.75 for the r filter. We then converted the power-law radial profile into a 2D circular model (middle panel of Fig. \ref{ima:fitg}) centered on the star position and finally subtracted it from the original image. The right panel of Fig. \ref{ima:fitg} shows the Gemini data after removal of the scattered light model.

\subsubsection{Scattered light removal around NGC1052-DF4}
\label{sub:PSFandngc1035}

In the case of NGC1052-DF4, there are no bright stars (m$_V$$<$9 mag) in close proximity (d$<$5\arcmin) to the galaxy as in the case of NGC1052-DF2.
Nevertheless, there are a few bright objects that could introduce unwanted scattered light and complicate the analysis of the outskirts of this galaxy.
One is the galaxy NGC1035 to the east (see Fig. \ref{fig:dithering-df4}). The other is the star at RA = 39.7779 and Dec = -8.1126, to the west of NGC1052-DF4. This star has a brightness of Gmag = 14 mag.

\begin{figure*}
    \centering
    \includegraphics[width=\textwidth]{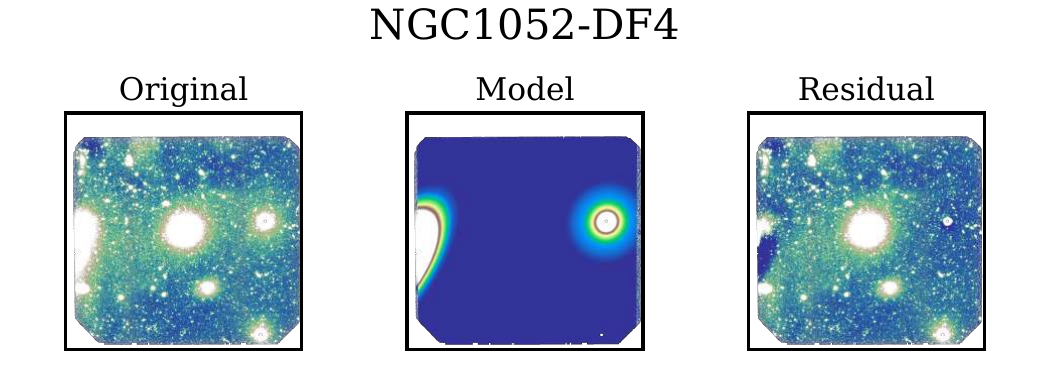}
    \caption{Three stages of subtraction of the scattered light emitted by NGC1035 and the star (Gmag = 14 mag) near NGC1052-DF4 (similar to Fig. \ref{ima:fitg}). The panels depict the mosaic of the field (5\arcmin$\times$5\arcmin) using a color image generated with r-filter Gemini data. The left panel shows the original stacked image, and the middle panel illustrates the models of the star and the galaxy NGC1035. The right panel shows the residual image generated by subtracting the models from the original image.}
    \label{fig:slremovaldf4}
\end{figure*}

To subtract both the nearby star and the galaxy, we created light models using \texttt{build-ellipse-model} in \texttt{Photutils}. 
The advantage of this method is the ability to change the PA and ellipticity of the isophotes at different radii to create more realistic models and minimize residuals. To obtain a satisfactory fit of NGC1035, whose physical center is outside the FoV of the Gemini camera, we used a technique using \texttt{Gnuastro}'s \texttt{astwarp} to combine the Gemini data with surrounding INT \citep{Montes2021} data. To prepare the data for modeling, we also applied a mask based on the \texttt{Noisechisel} detection map. Then, by deriving the optimal models for the star and galaxy, we performed the subtraction of these models (middle panel of Fig. \ref{fig:slremovaldf4}) from the original data. The right panel of Fig. \ref{fig:slremovaldf4} shows the Gemini data after removal of the
scattered light models.

\subsection{Surface brightness limits of the final co-adds}
\label{sub:sblimits}

\begin{table*}
    
    \centering
    \begin{threeparttable}
        \caption{Surface brightness limiting depths for the different datasets analyzed in this paper.}
        \label{tablelimits}
        \begin{tabular}{c|c|c|c|c|c}
            \hline
            \hline
            \multicolumn{6}{c}{Surface brightness limiting depth (3$\sigma$; 10\arcsec $\times$ 10\arcsec\ boxes)} \\
            \hline
            \hline
            \multicolumn{6}{c}{NGC1052-DF2} \\
            \hline
            Filter & SDSS\tnote{a} & DECaLS\tnote{a} & Dragonfly\tnote{b} & INT\tnote{c} & Gemini\tnote{a} \\
            \hline
            r-Sloan & 26.9 mag/arcsec$^2$ & 28.7 mag/arcsec$^2$ & -- & 29.5 mag/arcsec$^2$ & 30.5 mag/arcsec$^2$ \\
            g-Sloan & 27.5 mag/arcsec$^2$ & 29.1 mag/arcsec$^2$ & 29.35 mag/arcsec$^2$ & 30.4 mag/arcsec$^2$ & 30.9 mag/arcsec$^2$ \\
            \hline
            \hline
            \multicolumn{6}{c}{NGC1052-DF4} \\
            \hline
            & SDSS\tnote{a} & DECaLS\tnote{a} & Dragonfly & IAC80\tnote{d} & Gemini\tnote{a} \\
            r-Sloan & 26.9 mag/arcsec$^2$ & 28.7 mag/arcsec$^2$ & -- & 29.3 mag/arcsec$^2$ & 30.6 mag/arcsec$^2$ \\
            \hline
        \end{tabular}
        \begin{tablenotes}
        \item[a] The limiting surface brightness has been derived  using the methodology  described in Appendix A of \cite{Javicirri2020}.
            \item[b] \cite{Keim2022}. The value given in the table corresponds to the limiting depth given in  https://arxiv.org/pdf/2109.09778v1 (see their Sect. 3.3).
            \item[c] \cite{Montes2021}.
            \item[d] \cite{Montes2020}.
        \end{tablenotes}
    \end{threeparttable}
\end{table*}

The NGC1052-DF4 and NGC1052-DF2 mosaics are generated using the data reduction pipeline described above. Each field has a different scattered light distribution that is processed individually. Once this was done, we estimated the surface brightness depth of our images. The surface brightness limits we obtained using the metric 3$\sigma$ fluctuations in 10" $\times$ 10" boxes are 30.9 mag/arcsec$^2$ (g band) and $\sim$30.5 mag/arcsec$^2$ (r band; see Appendix A of \cite{Javicirri2020} for more details on the implementation of the chosen metric). To put our new images in context with  previous datasets, in Table \ref{tablelimits} we show the limiting surface brightness of earlier images using the same metric as in the work presented here. With the exception of the g-band data presented in \citet{Montes2021}, the Gemini images are at least 1 mag deeper than previous works.

\section{Analysis}
\label{sec:analysis} 

In this section we use our ultra-deep images to check for the presence of tidal features and to analyze the light distribution around NGC1052-DF2 and NGC1052-DF4.

\begin{figure*}
    \centering
    \includegraphics[width=\textwidth]{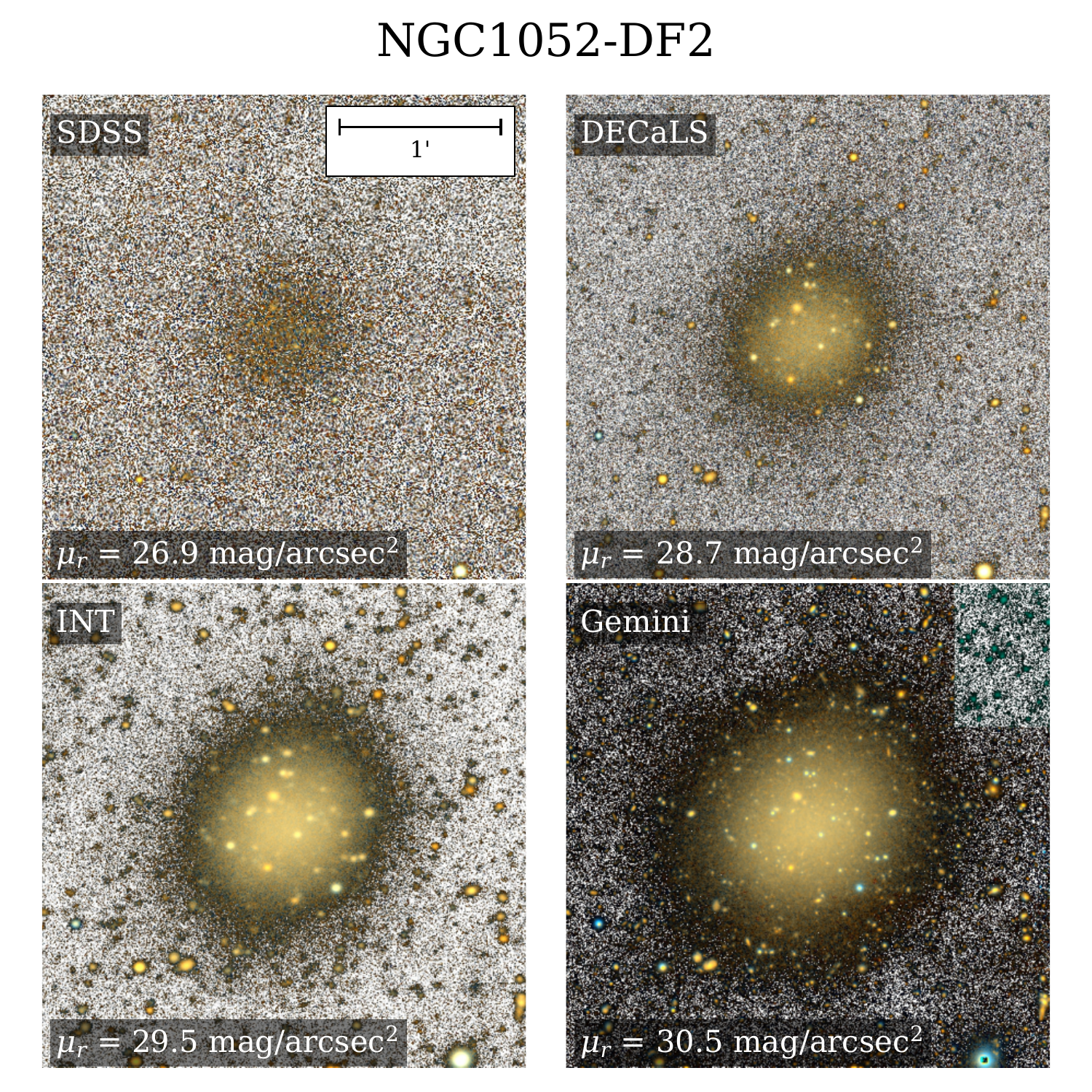}
    \caption{Appearance of NGC1052-DF2 at different surface brightness depths. The FoV of each stamp is 3\arcmin\ $\times$ 3\arcmin\ . The limiting surface brightness of the r-band images (SDSS, DECaLS, INT, and Gemini) is indicated on the panels using a common metric of 3$\sigma$ in 10\arcsec $\times$ 10\arcsec\ boxes. The color of the images is a combination of the g and r filters. The black and white background is from the g-band filter. As the image gets deeper, the apparent size of the galaxy also increases, with no obvious signatures of distortions in its periphery.}
    \label{fig:visual-comp-df2}
\end{figure*}

\subsection{NGC1052-DF2}
\label{sub:df2}

To put the depth of the current dataset into context, Fig. \ref{fig:visual-comp-df2} shows a region of 3\arcmin$\times$3\arcmin\ around NGC1052-DF2 using different datasets, namely SDSS, DECaLS (DR10), INT, and Gemini. The limiting surface brightness of each dataset is shown in the figure.
The color of the images is constructed using \textit{astscript-color-faint-gray} \citep{scriptcolorimagesRaul}, by combining the g and r filters, with the white and black background corresponding to the g filter.
The apparent size of the galaxy increases with image depth.


\subsubsection{Radial surface brightness profiles}
\label{subsub:sbdf2}

\begin{figure}
    \centering
    \includegraphics[width=0.5\textwidth]{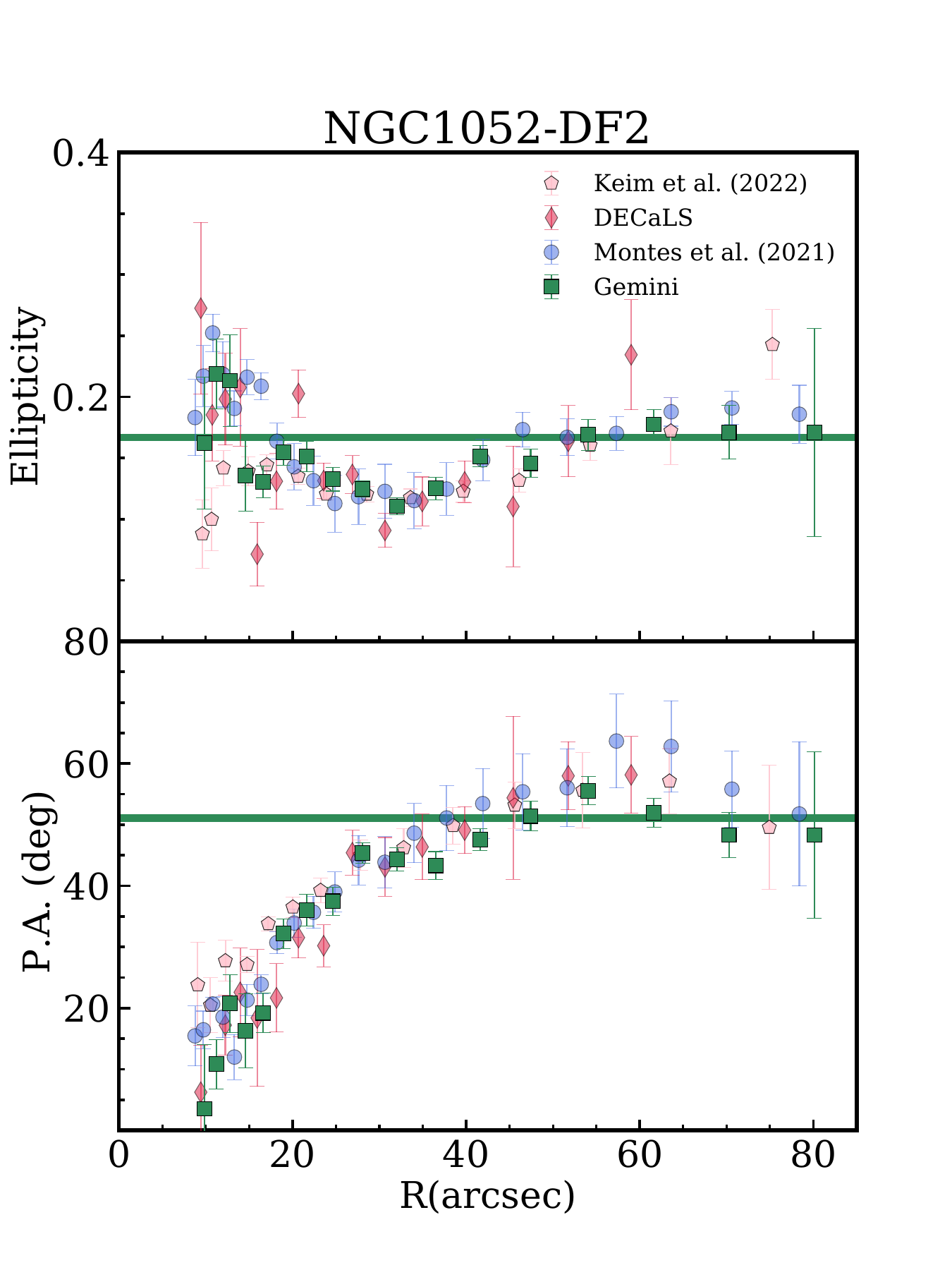}
    \caption{Ellipticity (top) and PA (bottom) radial profiles of NGC1052-DF2 using different datasets. The profiles for the DECaLS and Gemini images were derived by our group using \texttt{ellipse}, starting the fit at 40\arcsec. The radial profiles for INT and Dragonfly are those published in \citet{Montes2021} and \citet{Keim2022}, respectively. The green horizontal lines correspond to the average ellipticity (top) and PA (bottom) values in the radial range between 40\arcsec\ and 80\arcsec.}
    \label{fig:pa-ell-df2}
\end{figure}


To derive the surface brightness profiles of NGC1052-DF2 in both bands, we first created a common mask for both filters using an image resulting from the sum of the g and r bands. The mask is created to avoid contamination from foreground and background sources. A first step in generating the mask is to run \texttt{Noisechisel} and \texttt{Segment}. As a second step, to improve the detection of the contamination on top of our galaxies, we applied the technique of unsharp masking to remove the light of the galaxy and thus easily detect the individual contamination sources on top of it. Unsharp masking consists of smoothing the original image with a Gaussian filter of the width of the point-like sources and subtracting the smoothed image from the original. Then we applied \texttt{Noisechisel} and \texttt{Segment} again to mask the remaining undetected objects.
The mask applied to NGC1052-DF2 field is shown in the left panel of Fig. \ref{ima:mask-df2df4}.


Once all sources of contamination are masked out, we ran \texttt{ellipse} \citep{Jedrzejewski1987} in \texttt{Photutils} \citep{Bradley2019} to obtain the best PA and ellipticity of the isophotes in the outer parts of the galaxy.
This code fits elliptical isophotes to the 2D images of galaxies using the method described in \citet{Jedrzejewski1987}. We first ran \texttt{ellipse} allowing all parameters to vary freely.  In the second run, we fixed the center to the median centers of the isophotes returned by \texttt{ellipse} in the first iteration\footnote{The uncertainty associated with the selection of the center of the galaxy is $\pm$ 0.3 \arcsec in both RA and Dec.}.
\texttt{ellipse} fits the isophotes both outward and inward from an initial semimajor axis (SMA) length.
In both cases, we started the fit with a radius of 40$\arcsec$ and used a linear distance between ellipses\footnote{Starting the fit at different locations (between 20\arcsec\ and 50\arcsec) results in slight variations of the ellipticity and PA of 0.01 and 3 deg, respectively. We selected 40\arcsec\ as better characterization of this variation.}. We report the results of the second run of the \texttt{ellipse} routine for NGC1052-DF2 (for each of the bands used in this paper) in Fig. \ref{fig:outputdf2}.

In Fig. \ref{fig:pa-ell-df2} we make a comparison of the ellipticity and PA radial profiles derived from our data with those found in the literature. This analysis is relevant because changes in ellipticity and PA with radius could indicate changes in the structure of the galaxy. These changes could be due either to internal structures (triaxiality, spiral arms, etc.) or to external effects (tidal distortions, minor mergers, etc.). To make such a comparison, we used the combined g+r image\footnote{The same analysis using each band separately is shown in in Fig. \ref{fig:outputdf2}.}. This was done to provide a consistent comparison with other analyses in previous papers that used this combined information. There is generally good agreement (within the error bars) between the different datasets. Beyond 40\arcsec\ \citep[equivalent to $\sim$2R$_e$ with R$_e$ = 22.6\arcsec;][]{discoverydf2} the ellipticity and PA of the galaxy are fairly constant. We estimated these values by fitting a constant to the values between 40\arcsec\ and 80\arcsec\ (green horizontal lines in Fig. \ref{fig:pa-ell-df2}). We find an axis ratio b/a= $\dftwoAR{}$$\pm0.01$ and PA= $\dftwoPA{}$$\pm$2.7 deg. Based on the elliptical and PA radial profiles, NGC1052-DF2 appears to have two separate structural components, before and after a radius of $\sim$40\arcsec. Since the image of the galaxy presented in Fig. \ref{fig:visual-comp-df2} shows no obvious signature of distortion, we favor the hypothesis that this structural change is most likely related to an internal change in the morphology of the galaxy. We discuss this idea in Sect. \ref{sec:discussion}.

Before extracting the surface brightness profile, we needed to correct our images for any residual sky contribution near the galaxy.
To do this, we computed the value of the background level of the data in a masked elliptical annulus centered on the galaxy. This elliptical annulus has the shape that is representative of the outer part of the galaxy. That is, it has the mean PA and mean axis ratio values we derived above. The annulus has inner and outer radii of 120$\arcsec$ and 140$\arcsec,$ respectively.

\begin{figure} 
    \centering
        \includegraphics[width=0.5\textwidth]{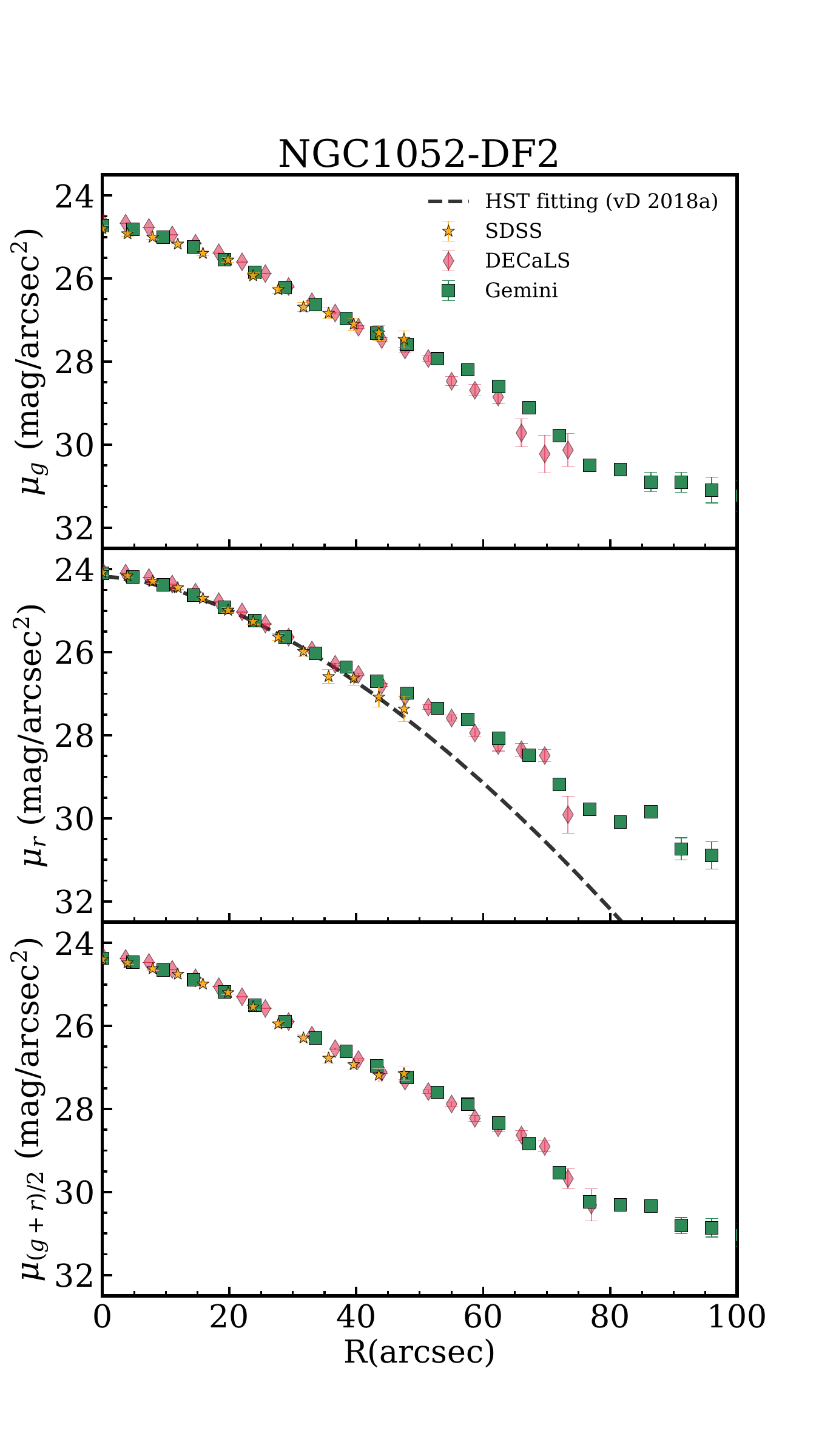}
        \caption{Surface brightness profiles of NGC1052-DF2 in the g band (upper panel), r band (middle panel), and the combination of the two filters (lower panel). The different symbols correspond to data obtained with different telescopes. These profiles were obtained using a fixed ellipticity and PA representative of the outer parts of the galaxy (see the main text for details). The dashed line in the middle panel corresponds to the S\'ersic fit to the F606W band of the HST ACS camera \citep[n = 0.6 and Re = 22.6$\arcsec$; ][]{discoverydf2}.}
        \label{fig:df2-sbprof}
\end{figure}

Once the local sky value is subtracted from the image, we extracted the radial surface brightness profiles of the g and r filters using \textit{astscript-radial-profile} \citep{astscriptradialprofileRaul}. To get a better representation of the outer part of the galaxy, we set the ellipticity and PA to the average values we found above. This is particularly useful if we want to extract information beyond R=80\arcsec\, where the uncertainties in measuring the ellipticity and PA of the outer isophotes increase significantly. Using the same ellipticity and PA values, we extracted the surface brightness profiles of the SDSS data and DECaLS data. For consistency, the mask that we applied to these shallower datasets is the one derived for the Gemini data, but scaled to the respective pixel sizes (0.396\arcsec/pix for SDSS and 0.262\arcsec/pix for DECaLS). The results of doing this are shown in Fig. \ref{fig:df2-sbprof}.


The errors in the surface brightness profiles are calculated taking into account the two main sources of uncertainty in the estimate of the intensity at a given radial distance. Such an intensity is the sum of the intensity of the source itself and the intensity of the background. In short, we want to calculate the error associated with the sum I$_{tot}(r)$=I$_{source}(r)$+I$_{background}$ using the pixels located at a given radial distance within an elliptical annulus aperture.

The intensity of the background in ground-based optical dark observations, as in our case, is mainly produced by the brightness of the sky atmosphere ($\mu$$_{V}$ $\sim$ 22 mag/arcsec$^2$). There are other sources of background light that could be relevant, such as zodiacal light and the scattered light produced by nearby bright sources. In the particular case of our observations, these other background sources are subdominant with respect to the brightness of the night sky emission. For example, the scattered light from the nearby star HD16873 is contributing a surface brightness of $\mu_{V}$ >28 mag/arcsec$^2$ to the surroundings of NGC1052-DF2.

For both the source and background intensities, the error is characterized by the shot noise, and therefore the uncertainty per pixel for each component could be written as$\sqrt{I_{source}}$ and $\sqrt{I_{background}}$. In practice, since we are dealing with many pixels within the elliptical aperture, the characterization of the noise of each component can be done with good accuracy by measuring the standard deviation of the counts of each component within the elliptical aperture using only the pixels that are not masked (i.e.,  $N_{pix,ellipse}(r)$). Thus, when measuring the intensity profile of the object, we can estimate $\sigma_{tot}(r)$ by calculating the standard deviation of the counts within such an elliptical aperture. Under the Gaussian approximation, the uncertainty E$_{tot,r}$ when measuring I$_{tot}$ is therefore

\begin{equation}
    \label{eq:errstot}
    E_{tot,r} = \frac{\sigma_{tot}(r)} {\sqrt{N_{pix,ellipse}(r)}}
.\end{equation}

To estimate the error associated with the determination of the background intensity, we can use the background pixels of the image to accurately determine $\sigma_{background}$, which, assuming the sky background intensity is a constant, is independent of the radial position of the object isophote we are analyzing. However, the uncertainty E$_{background,r}$ in characterizing the contribution of I$_{background}$ to I$_{tot}(r)$ depends on the number of pixels within the elliptical aperture. For this reason, we estimated the E$_{background,r}$ when measuring the contribution of I$_{background}$ at each radial distance as
\begin{equation}
    \label{eq:errsky}
    E_{background,r} = \frac{\sigma_{background}}{\sqrt{N_{pix,ellipse}(r)}}
.\end{equation}Having measured the uncertainty at E$_{tot,r}$ and E$_{background,r}$, we can then estimate the uncertainty associated with the source itself, E$_{source,r}$:

\begin{equation}
    E_{source,r} = \sqrt{E_{tot,r}^2 + E_{background,r}^2}
    \label{eq:errtot}
.\end{equation}

In Fig. \ref{fig:df2-sbprof} we compare the surface brightness profiles in both filters (and their combination) obtained from our ultra-deep Gemini images with those derived using the SDSS and DECaLS datasets. Here, the surface brightness profiles correspond to those derived using fixed ellipticity and PA. There is a notable change in the surface brightness slope around 60\arcsec. This sudden decrease (also called truncation) has been observed in other dwarf galaxies with similar stellar masses \citep{2022nushkia}. The fact that the truncation is visible in both bands simultaneously, and also in the DECaLS data, gives us confidence that this feature is likely real. 
The surface brightness profiles when the ellipticity and PA are allowed to vary are shown in Fig. \ref{fig:sb-comparison}. Interestingly, the feature is less obvious when the ellipticity and PA are allowed to vary when deriving the surface brightness profiles (see Fig. \ref{fig:sb-comparison}). 
The significance of this is discussed in Appendix \ref{app:profiles}. We also include in the middle panel of Fig. \ref{fig:df2-sbprof} the S\'ersic fit to the \textit{Hubble} Space Telescope (HST) F606W band of the HST ACS camera made to the galaxy in \citet{discoverydf2}. This fit is a good representation of the light distribution of the galaxy in the innermost region, but differs from what has been observed with deeper data beyond 40\arcsec.

Finally, taking advantage of our very deep dataset, we derived the effective radius using the surface brightness profiles in the two different filters.
By integrating the light profiles extracted using fixed ellipses, we obtain R$_{eff,g}$ =25.8\arcsec$\pm$0.07\arcsec and R$_{eff,r}$ = 25.6\arcsec$\pm$0.07\arcsec. If the ellipticity and PA of the ellipses from which the surface brightness profiles are obtained are left free, the effective radii grow a bit:  R$_{eff,g}$ = 28.3\arcsec $\pm$ 0.17\arcsec and R$_{eff,r}$ = 28.6\arcsec $\pm$ 0.19\arcsec. This is as expected, since there is more light in the outer parts of the galaxy (see Fig. \ref{fig:sb-comparison}).   In this work, using fixed ellipses, for SDSS we get R$_{eff,g}$ = 21.5\arcsec $\pm$ 1\arcsec\ and R$_{eff,r}$ = 22.1\arcsec$\pm$1\arcsec, while for DECaLS we get R$_{eff,g}$ = 23.5\arcsec$\pm$0.2\arcsec and R$_{eff,r}$ = 24.1\arcsec$\pm$0.2\arcsec. In these cases the effective radii are lower because the data are shallower and therefore the light beyond a given radius is not measured. The Gemini effective radii are also comparable to those published in previous papers. For example, using HST data, \citet{discoverydf2}
finds R$_{eff,HST}$ = 22.6\arcsec\ (F606W; similar to the r band). This lower value is reasonable considering that the outer part of the galaxy is brighter than originally thought by fitting a S\'ersic law to the central region (see Fig. \ref{fig:df2-sbprof}, middle panel). With INT deep data, \citet{Montes2021} measure 28.4\arcsec$\pm$0.3\arcsec using fixed ellipses, in very good agreement with our measurements when ellipticity and PA are left free. Finally, using Dragonfly, \citet{Keim2022} derives R$_{eff,DF2}$ = 24.8\arcsec. This last measurement is obtained by a S\'ersic fit with n = 0.6, so it may be missing the light in the outer parts of the galaxy.

\subsubsection{Color and surface mass density profiles}
\label{subsub:color}

\begin{figure*}
    \centering
    \includegraphics[width=\textwidth]{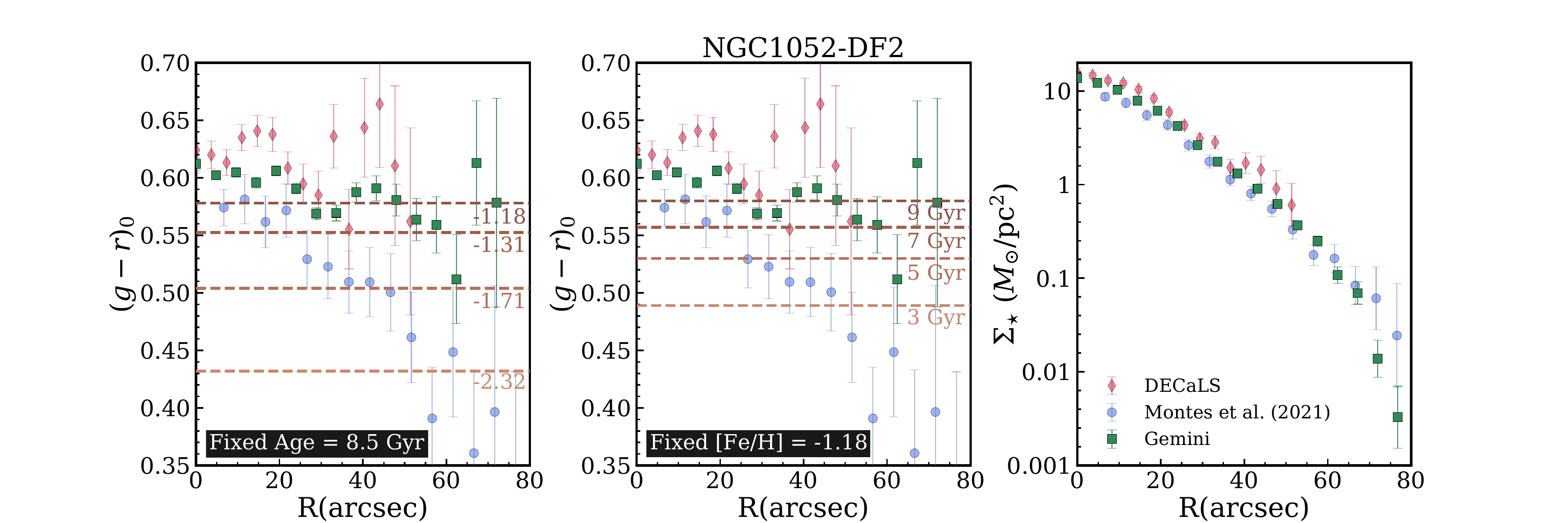}
    \caption{Color (g-r)$_0$ and stellar surface mass density profiles of NGC1052-DF2 using data from Gemini, INT \citep{Montes2021}, and DECaLS. In the left and middle panels, the horizontal dashed lines represent the expected color values based on the \cite{vazdekis} models for different metallicities at a fixed age of 8.5 Gyr (left panel) and for different ages at a fixed [Fe/H] of -1.18 (middle panel). The right panel shows the stellar mass density profile of NGC1052-DF2.}
    \label{fig:color-age-metallicity}
\end{figure*}

To better understand the nature of NGC1052-DF2, we examined its color g-r radial profile and its stellar surface mass density profile. This is shown in Fig. \ref{fig:color-age-metallicity}. The left and middle panels of Fig. \ref{fig:color-age-metallicity} show the g-r color profile of NGC1052-DF2 as a function of radial distance from the galaxy center for Gemini, INT, and DECaLS data. The color profile has been corrected of Galactic extinction using the following coefficients:  $A_{g}$ = 0.08 and $A_{r}$ = 0.06 \citep[][]{2011Schlafly}. In the case of Gemini and DECaLS data, the ellipticity and PA are fixed as explained above. The INT profile corresponds to the one published in \citet{Montes2021} and was also obtained using fixed ellipses. The INT color profile agrees on a bluish behavior toward the edge of the galaxy. This is not so clear in the DECaLS data. The Gemini profile seems to lie between these two shallower datasets. The small offsets ($<$0.02 mag) between the different datasets correspond to possible uncertainties in the calibration of the different bands. We must remember that neither the filters nor their response are exactly the same across the different facilities, so small changes could be expected. The error bars of the color profile are obtained by summing in quadrature the errors on the surface brightness profiles in the g and r bands.

With the goal of interpreting the meaning of the radial color profile of NGC1052-DF2, we explored what implications it might have for understanding the age and metallicity variations along the structure of the galaxy. We explored two scenarios. In the first (left panel), the age of the stellar population is fixed at 8.5 Gyr and we plot the expected color as a function of metallicity (dashed lines) using MILES models \citep{vazdekis}. In the middle panel we repeat the procedure, but this time fixing the metallicity at [Fe/H] = -1.18 and plotting the expectations for different ages. The age and metallicity chosen in this exercise are motivated by the values found in the center of the galaxy using spectroscopy \citep[see, e.g.,][]{2019MNRAS.486.5670R}. The Gemini data are consistent with a moderate decrease in the age and/or metallicity of the galaxy toward the outskirts.
We performed a linear fit of the form (g-r)$_0$(R) = m$\times$R + c to the Gemini radial color profile. We find m = -(8.5$\pm$1.2)$\times$10$^{-4}$ and c = 0.61$\pm$0.03. The null hypothesis (i.e., m = 0) is rejected at 7$\sigma$.

This analysis suggests that the observed color gradient in the galaxy could be: an age gradient, a variation in stellar metallicity, or a combination of both. However, as previously pointed out by \cite{Montes2021}, it is difficult to distinguish between age and metallicity variations based on a single color.
Nevertheless, the lack of HI detection in this galaxy \citep{2019MNRAS.482L..99C} suggests that the galaxy may have a uniform age throughout its structure. In addition, previous observations of MUSE data of NGC 1052-DF2 by \cite{2019fensch} indicate a possible metallicity gradient.
Assuming a uniform age for the entire galaxy, the observed color gradient implies a possible increase in metal-poor stellar populations toward the outer regions.

Finally, in the right panel of Fig. \ref{fig:color-age-metallicity} we show the stellar surface mass density radial profile of NGC1052-DF2. To derive this profile we used the prescriptions given in \cite{Bakos2008}. We used the g-band profile as the reference band, and we estimated the (M/L)$_g$ at each radial distance using the g-r profile (extinction corrected) using \cite{2015MNRAS.452.3209R}. We assumed a \cite{chabrier2003} initial mass function (IMF). Changing the IMF will scale the stellar mass density profile up and down, affecting the total stellar mass, but will not change the global shape of the profile. In addition to the Gemini data, we also show the stellar surface mass density using the DECaLS data (fixed ellipses, same as Gemini) and the INT data \citep[fixed ellipses with PA and ellipticity found in][]{Montes2021}. The slight vertical offsets between the different profiles correspond to the small offsets in the g-r color that produce different (M/L)$_g$. In general, there is a nice agreement up to 70\arcsec. Beyond this radius, the Gemini profile shows a strong indication of a truncation (edge) in the stellar mass profile. This is not so obvious in the case of the INT data. The reason for this is the shape of the surface brightness profile in the g band in the case of the INT data. As shown in \citet{Montes2021}, the outer part of NGC1052-DF2 is brighter in this band than what we find in the Gemini data. This is not observed in the r band where both works agree very well on the surface brightness profiles. The excess light in the INT g-band data is also responsible for the strong bluing in the g-r profile for that data. Interestingly, if the clear truncation on the Gemini profile is correct, the stellar surface mass density where it happens (i.e., $\sim$0.1 M$_{\sun}$/pc$^2$) is within the expectation for the values found for the location of the truncation by \citet{2022nushkia} (see their Fig. 5) for galaxies of similar stellar mass (i.e., $\sim$10$^8$ M$_{\sun}$).

\subsection{NGC1052-DF4}
\label{sec:df4}

\begin{figure*}
    \centering
    \includegraphics[width=17cm]{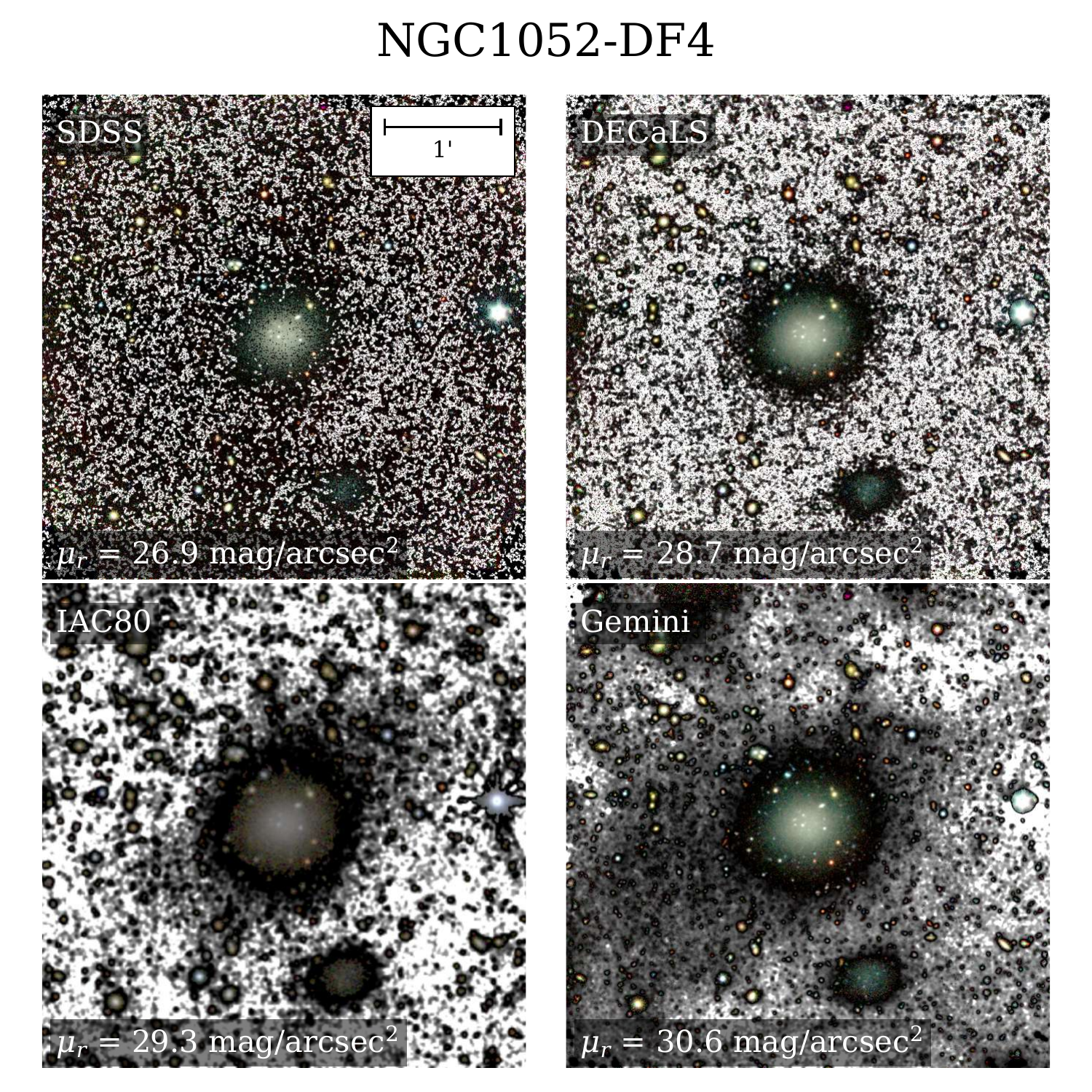}
    \caption{NGC1052-DF4 appearance as the image depth increases. The four stamps presented, each covering a 4\arcmin$\times$4\arcmin\ region around the galaxy NGC1052-DF4, correspond to observations obtained at different surface brightness limiting depths in the r band (3$\sigma$, 10\arcsec$\times$10\arcsec) using various facilities. The tidal streams become visible when reaching surface brightness levels fainter than $\mu_{r}$$\sim$29 mag/arcsec$^2$. In all stamps, to represent different pixel intensities, the white corresponds to faint pixels and black represents brighter features. The black and white background is generated used the r band only. The color part of the stamps is created using rescaled HiPERCAM data from \citet{Montes2020} using the g, r, and z filters.}
    \label{fig:visual-comp-df4}
\end{figure*}

Figure \ref{fig:visual-comp-df4} illustrates the gradual emergence of tidal features around NGC1052-DF4 as the depth of the data increases. The four stamps shown, each covering a square region of 4\arcmin$\times$4\arcmin\ around the galaxy NGC1052-DF4, correspond to observations made at different surface brightness limiting depths using a variety of facilities. While the presence of tidal streams remains invisible in the case of the SDSS and DECaLS, both the IAC80 and Gemini data show a clear excess of light in the outskirts of the object that we identify as tidal tails.


\subsubsection{Radial surface brightness profiles}
\label{sub:sbdf4}

To obtain the surface brightness profile of NGC1052-DF4, we followed the same strategy as for NGC1052-DF2. First, we carefully masked the sources of contamination that we find in the field. We used exactly the same tools and codes explained before.
The mask applied to the data is shown in the right panel of Fig. \ref{ima:mask-df2df4}.

\begin{figure}
    \centering
    \includegraphics[width=0.5\textwidth]{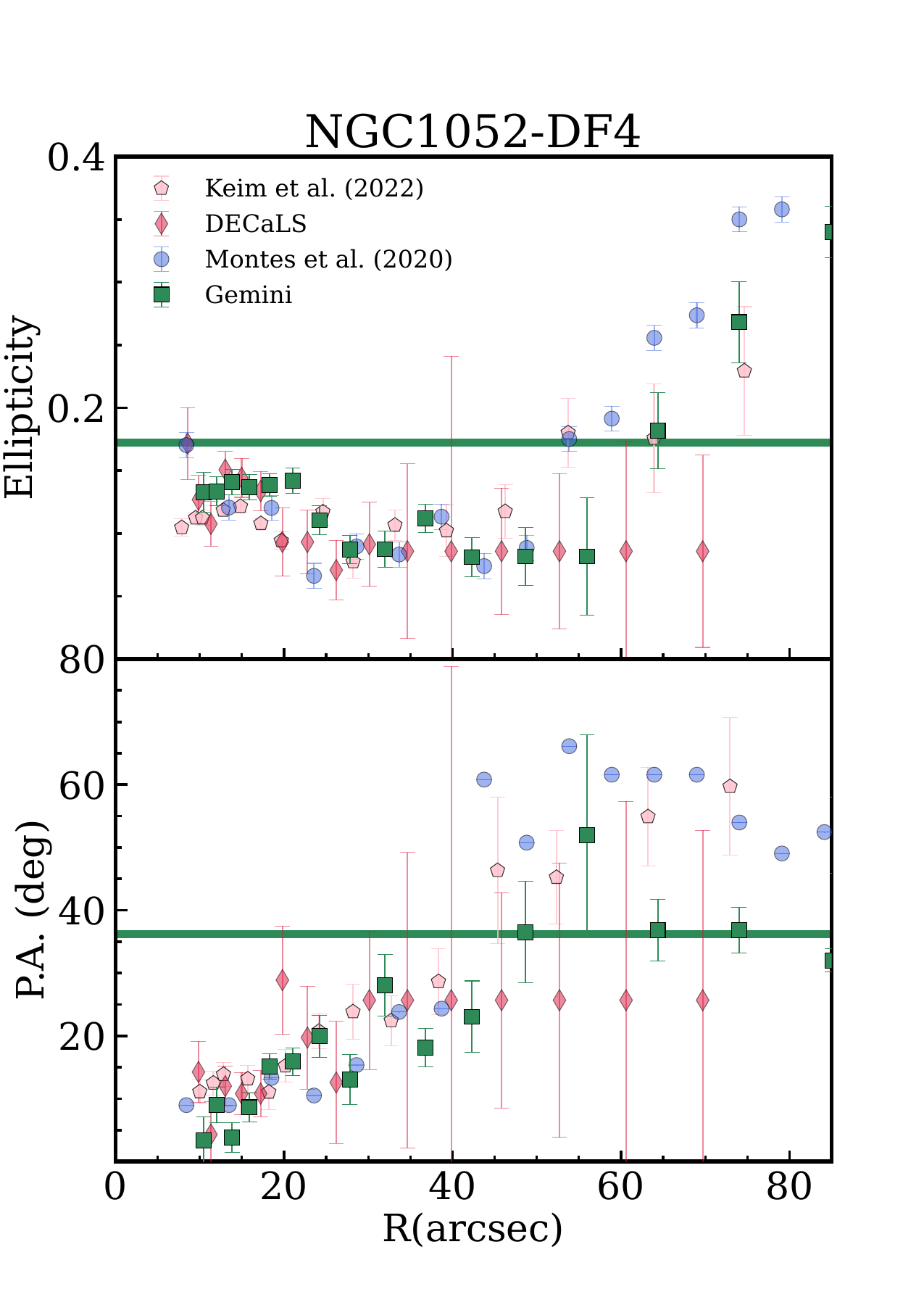}
    \caption{Ellipticity (top) and PA (bottom) radial profiles of NGC1052-DF4 using different datasets. The profiles for the DECaLS and Gemini images were derived by our group using \texttt{ellipse}, starting the fit at 30\arcsec. The radial profiles for IAC80 and Dragonfly are those published in \citet{Montes2020} and \citet{Keim2022}, respectively. The green horizontal lines correspond to a fit to the average ellipticity (top) and PA (bottom) values in the radial range between 40\arcsec\ and 80\arcsec.}
    \label{fig:pa-ell-def-df4}
\end{figure}

In the case of NGC1052-DF4, it is particularly important to be careful when determining the background level, because the tidal tails of the galaxy cover a significant fraction of the pixels of the image. For this reason, only circular regions (with radius equal to 10\arcsec) in the lower part of the image are used for this purpose. Next, we ran \texttt{ellipse} to derive the ellipticity and PA radial profiles. Our starting radius for fitting the shape of the isophotes is 30\arcsec. 
In Fig. \ref{fig:pa-ell-def-df4} we compare these ellipticity and PA profiles with those obtained from other images (DECaLS in this work) and from the literature: IAC80 \citep{Montes2020} and Dragonfly \citep{Keim2022}. The latter are the published ones and therefore correspond to those obtained from the sum of the g+r+i and g+r images, respectively. In the case of Gemini and DECaLS, we used only the r band. There is a reasonable agreement between the different datasets on the radial profile of the ellipticity. All the works agree on a sudden increase in this parameter at about 45\arcsec\, following the shape of the tidal tails of the galaxy (see the left panel of Fig. \ref{ima:outputdf4}).
The inner parts of the galaxy (R$<$45\arcsec) show two components for the ellipticity, one around 0.15 for R < 20\arcsec and then a drop to ellipticity = 0.1.

For the PA profile, the agreement between the different telescopes is less clear. All data agree on a slight increase in the PA up to R$\sim$40\arcsec. Beyond this radius, while all the works converge on an increase in the PA, the significance of this change depends on the dataset. We think that the differences between the different images could be related to the removal of the contamination caused by the nearest star and NGC1035. At these very low surface brightness levels, the size of the residuals after subtracting the light from these objects is crucial. Fortunately, as we can see in Fig. \ref{ima:sb-comparison-df4}, the effect on the shape of the surface brightness profile is not large and does not affect the interpretation of the tidal tails of the galaxy.

\begin{figure}
    \centering
    \includegraphics[width=0.5\textwidth]{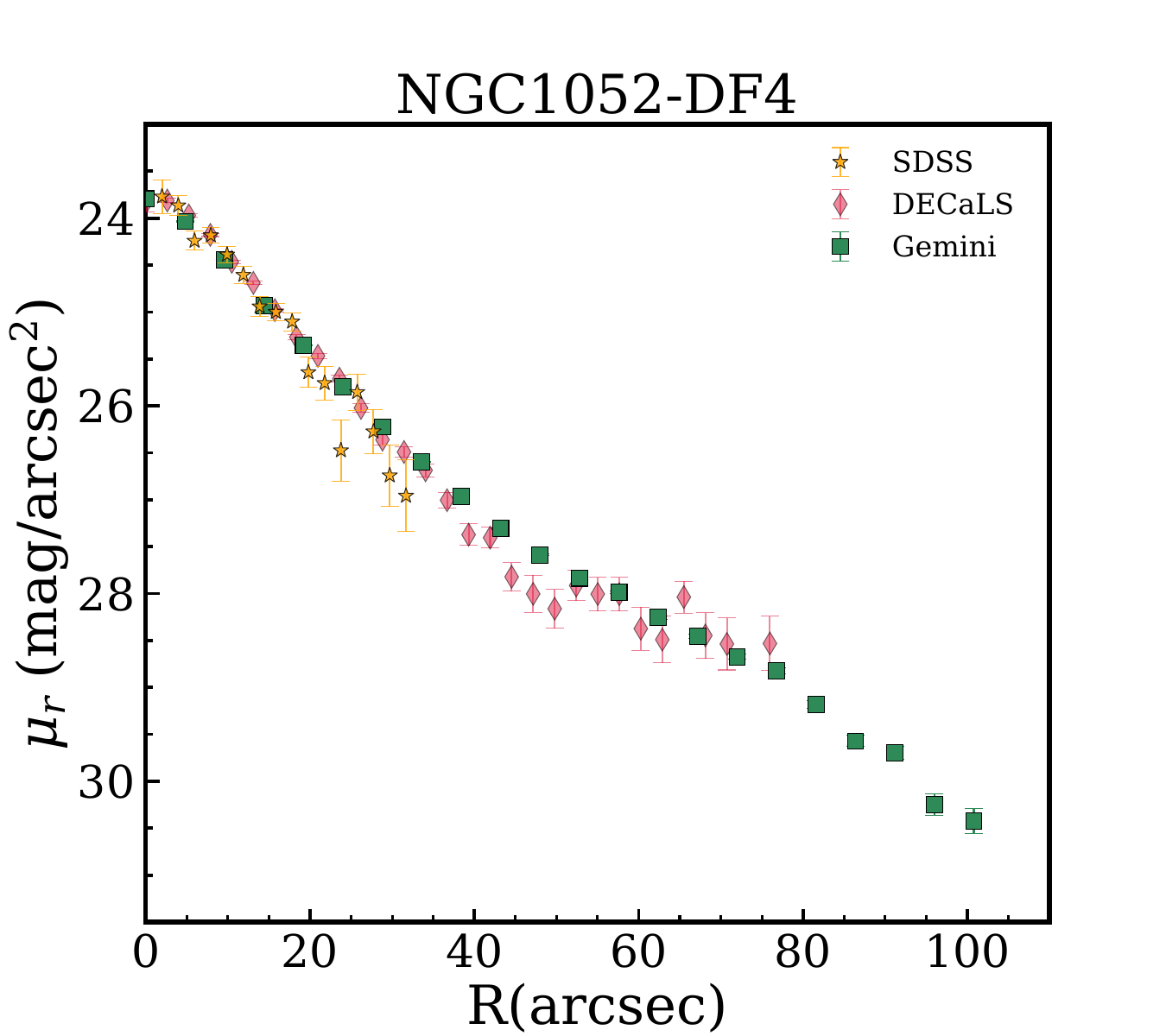}
    \caption{Surface brightness profiles of NGC1052-DF4 in the Sloan r band. The different symbols correspond to data obtained with different telescopes. These profiles were all obtained using a fixed ellipticity and PA (based on the Gemini data)  representative of the outer part of the galaxy (see the main text for details).}
    \label{fig:df4-prof}
\end{figure}

Following the same strategy as for NGC1052-DF2, we measured a representative value of axis ratio and PA for the outer isophotes of NGC1052-DF4 (40\arcsec$<$R$<$80\arcsec; see Fig. \ref{fig:pa-ell-def-df4}). The values we find are: b/a=\dffourAR{}$\pm$0.05 and PA=\dffourPA{}$\pm$4 deg. With these values fixed, we extracted the surface brightness radial profile of the galaxy. This is shown in Fig. \ref{fig:df4-prof}. Beyond R=40\arcsec, and in parallel with the increase in the ellipticity of the isophotes that we found earlier associated with the tidal tails of the galaxy, the surface brightness profile of the galaxy changes its slope, showing an excess of flux (an upward bend) with respect to the exponential decrease in the innermost region of the galaxy. This transition occurs at a surface brightness of about 27.5 mag/arcsec$^2$ (r band). To compare the profile with previously published data where the ellipticity and PA of the isophotes are left free, we show in Fig. \ref{ima:sb-comparison-df4} the surface brightness profile of NGC1052-DF4 using Gemini data and isophotes free. The results are qualitatively the same.

Very deep imaging of this system also allows us to study the effect of the appearance of the tidal tails on the measurement of the effective radius of the galaxy. Using the very deep Gemini data, the effective radius derived by integrating the light profile of the galaxy in the r band is: R$_{eff,r}$ = 24.5\arcsec$\pm$0.1\arcsec (using fixed ellipses) and R$_{eff,r}$ = 25.8\arcsec$\pm$0.1\arcsec (free ellipses). This slight increase in the effective radius is as expected, since the free ellipses better follow the light distribution in the outer parts of the object. In shallower data such as SDSS, the effective radius is significantly smaller (R$_{eff,r}$ = 15.3\arcsec$\pm$1.3\arcsec) because the data are not sensitive to the presence of tidal tails. For DECaLS we get R$_{eff,r}$ = 21.5\arcsec$\pm$0.5\arcsec. With respect to the values found in the literature, using Dragonfly, \citet{Keim2022} derived R$_{eff,DF4}$ = 19.8\arcsec\ by applying a S\'ersic fit with n=0.85, while using IAC80 deep data, \citet{Montes2020} obtained 18\arcsec\  by applying a S\'ersic fit with n=0.86. In both cases, while the fit does a good job of describing the inner parts of the galaxy (R$<$40\arcsec), it falls short to characterizing the outer part of the object affected by the tidal tails.


\section{Discussion}
\label{sec:discussion}

As we mentioned in the Introduction, ultra-deep imaging of both NGC1052-DF2 and NGC1052-DF4 is expected to greatly constrain their formation scenario and shed light on their low dark matter content. In the following, we discuss what the data obtained in this work, in addition to the information obtained in previous works, tell us about the nature of these objects. 

\subsection{NGC1052-DF2: No signatures of tidal disruption}

All the data we analyzed and collected from the literature agree very well with the ellipticity and PA radial profiles of NGC1052-DF2. The ellipticity of the object varies only slightly throughout its structure, with values fluctuating between 0.1 and 0.2. The PA does vary along the galaxy's structure, but beyond R=40\arcsec\ it remains fairly constant around 50 degrees.  These results on ellipticity and PA variation were already known and discussed in the literature. Despite the remarkable agreement between the different datasets, the interpretation is very different depending on the research group. For \citet{Montes2021}, the radial profile of the PA can be interpreted as a transition from a pressure-supported region (R$<$40\arcsec) to a disk-like outer part. For \citet{Keim2022}, however, the change in the PA plus the change in the surface brightness profile (with transitions from a low S\'ersic index n$<$0.6 to an exponential shape) is associated with a tidal disturbance.

The exponential decrease in the radial surface brightness profile beyond R$>$40\arcsec\ is not the only argument that \citet{Montes2021} use to support the idea of a rotationally dominated structure in the outer part of NGC1052-DF2. In fact, these authors argue that the dynamics of the GC system of this galaxy is also consistent with a rotating disk \citep{lewis2020}.

The new Gemini ultra-deep imaging is useful to further reinforce or reject the two contending scenarios (i.e., rotating disk vs. tidal disturbance). In this work we find no signatures that support NGC1052-DF2 being tidally disrupted. It should be emphasized that the change in PA occurs at a relatively bright surface brightness ($\mu_g$$\sim$27 mag/arcsec$^2$). If this feature was related to the development of tidal tails in the galaxy, the Gemini data should  show the S-shaped structure that we can  see in NGC1052-DF4, for example. This is not the case. While it is arguable that the surface brightness profile of \citet{Keim2022} shows an excess of brightness in the outer part compared to the rest of the data (see, e.g., Fig. \ref{fig:sb-comparison}), this excess can be understood as the result of a different sky background subtraction. The background subtraction used in the present work, in \citet{Montes2021}, as well as in \cite{Keim2022} (see their Appendix B) is done by removing a constant value. Therefore, it cannot destroy a possible S-like shape around NGC1052-DF2.

\begin{figure}
    \centering
    \includegraphics[width=0.5\textwidth]{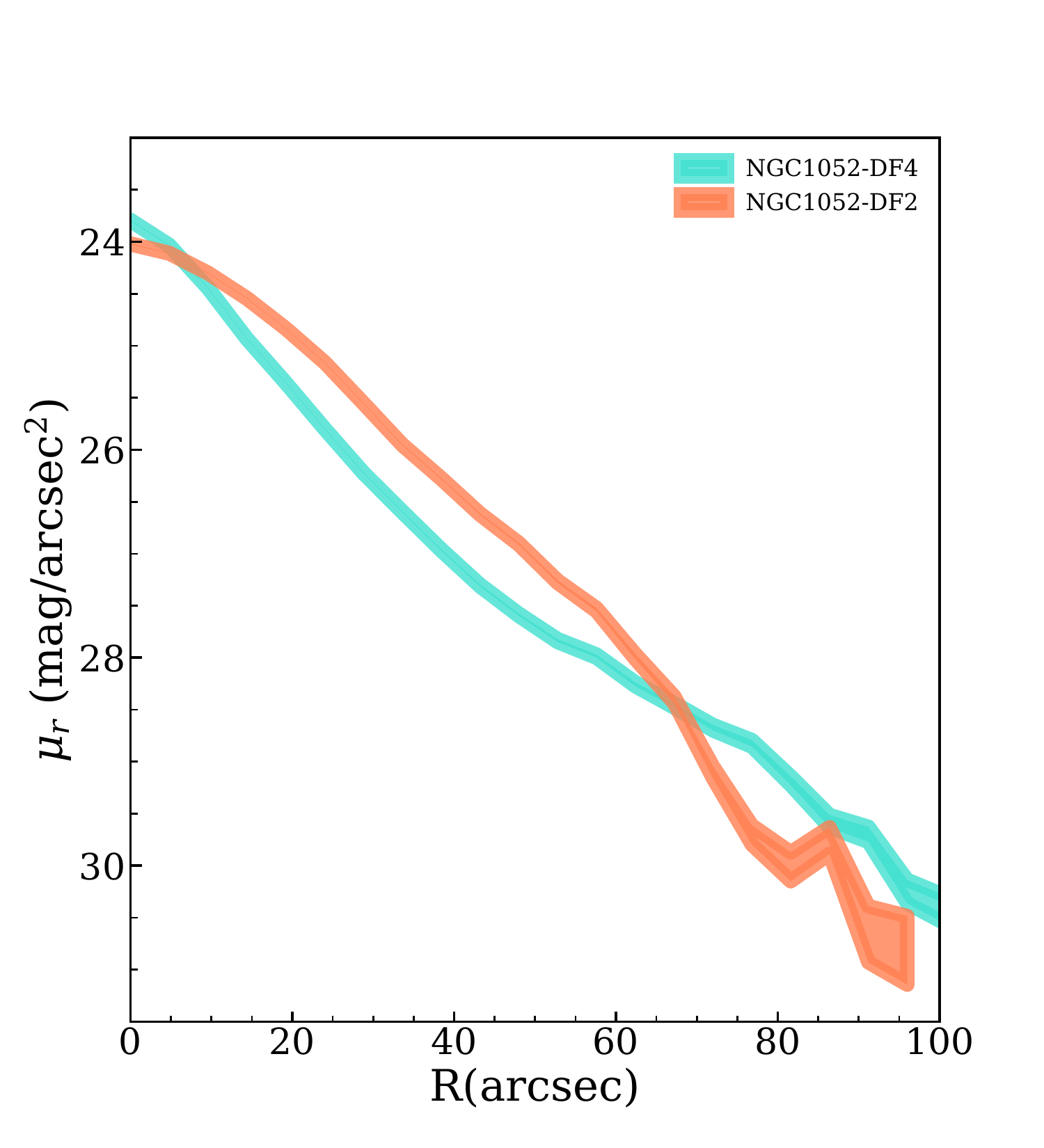}
    \caption{Surface brightness profiles in the r band of NGC1052-DF4 (light blue) and NGC1052-DF2 (orange) obtained from fixed ellipses with Gemini ultra-deep imaging. The shaded regions correspond to the error bars associated with the measurements. The two profiles are strikingly different. NGC1052-DF4, which is clearly tidally disturbed, shows an excess of light (R$>$40\arcsec) that is not seen in the case of NGC1052-DF2.}
    \label{fig:df2df4}
\end{figure}

Another way to look at this is to make a direct comparison between the surface brightness profile of NGC1052-DF2 (with no signatures of tidal distortions) and NGC1052-DF4 (with obvious distortions). This is done in Fig.\ref{fig:df2df4}. The two profiles are very different. NGC1052-DF2 shows an exponentially decreasing surface brightness profile up to R = 60\arcsec, with a hint of a break in the profile at R=60\arcsec. In contrast, NGC1052-DF4 shows an excess of light, an upward bend, in the outer regions (R$>$40\arcsec) of the surface brightness profile.

The absence of tidal features around NGC1052-DF2 up to the surface brightness limits of the image ($\mu_{g}$$\sim$31 mag/arcsec$^2$; 3$\sigma$; $10\arcsec \times 10\arcsec$ boxes) is useful to rule out a number of formation scenarios for this galaxy. For example, either an ongoing \citep{2023KATAYAMA} or a close flyby \citep{2022Moreno}  of NGC1052-DF2 around NGC1052 should have produced visible signatures of the interaction in the outskirts of this galaxy in images as deep as the ones we have here. Therefore, the likelihood that either of these scenarios has occurred should be low. The absence of tidal distortions is also useful to place strong constraints on the possible association of NGC1052-DF2 with its closer (in projection) massive galaxies: NGC1052 and NGC1042. This has been done in detail by \citet{Montes2021}. According to these authors, under the hypothesis that NGC1052-DF2 is physically connected to NGC1052, it should suffer from a tidal field that should have produced distortions well inside the galaxy body (55\arcsec$\pm$15\arcsec). This is rejected by the observations. However, if the galaxy is associated with NGC1042, no signatures of tidal distortions are expected down to $\sim$150\arcsec. This last scenario cannot be ruled out even with this deeper dataset. 

\begin{figure*}
    \centering
    \includegraphics[width=0.85\textwidth]{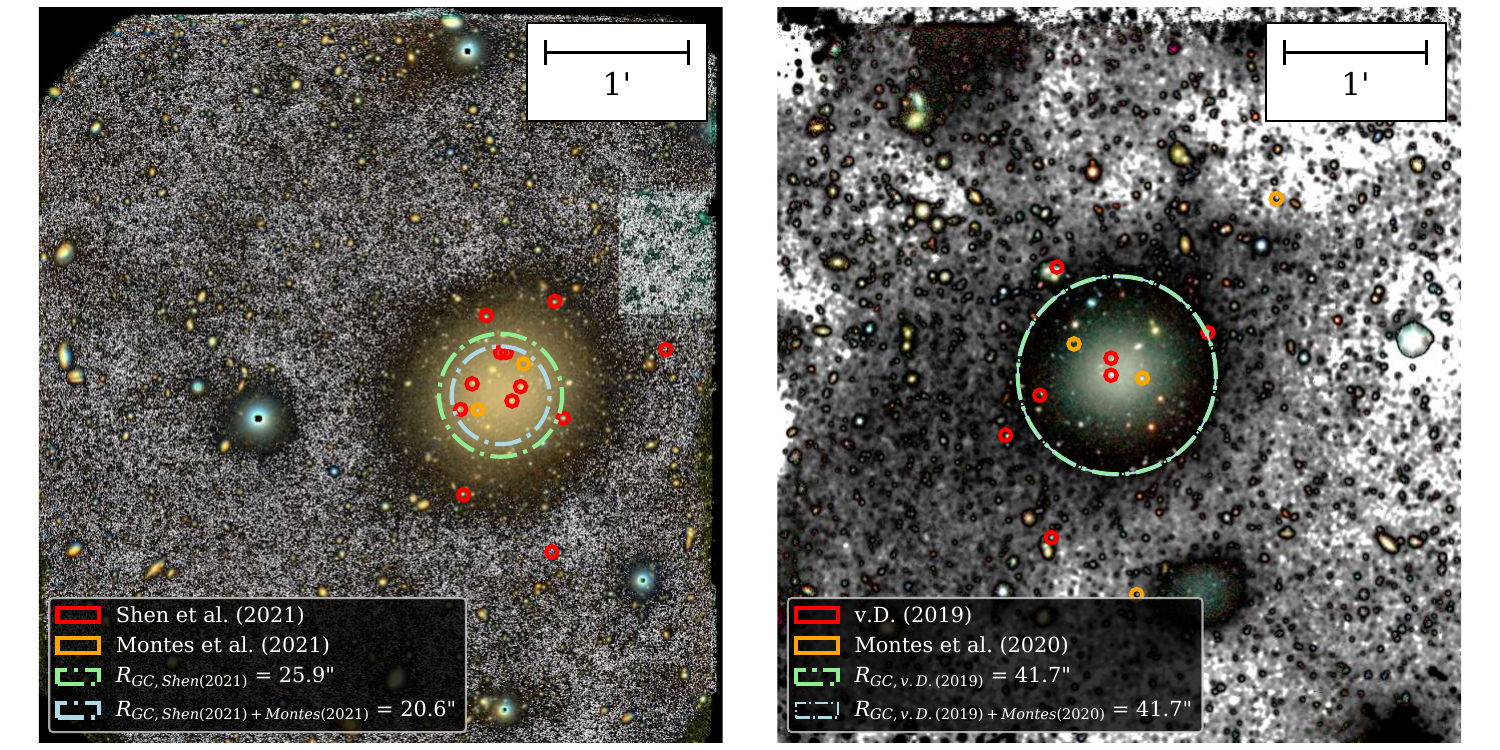}
    \caption{GC distributions around NGC1052-DF2 (left panel) and NGC1052-DF4 (right panel). Red circles indicate the positions of the spectroscopically confirmed GCs \citep{discoverydf4,2021shenA}, while the positions of the GC candidates found in \citet{Montes2020,Montes2021} are denoted with orange circles. The radii of the green circles (R$_{GC}$) represent the spatial location where half of spectroscopically confirmed GCs are found. The radii of the light blue circles instead, represent the spatial location where half of all GCs candidates are found.}
    \label{fig:globularclusters}
\end{figure*}

Another alternative way of testing the tidal disruption scenario for NGC1052-DF2 is by looking at the spatial distribution of its GC system. The object has eleven spectroscopically confirmed GCs \citep{2021shenA} with an additional four proposed in \cite{Montes2021}. 
In Fig. \ref{fig:globularclusters} (left panel) we show the distribution of the GCs orbiting NGC1052-DF2. It is remarkable that except for two, the vast majority of the GCs are directly on top the body of the object. This is suggestive that the object has not undergone a tidal disruption process. One way of quantifying this statement is by comparing the effective radius of the galaxy with the radius enclosing half of the GCs of the system (R$_{GC}$).

We determined R$_{GC}$ by computing the median of the   distances of the GCs in logarithmic units. If we consider only the spectroscopically confirmed GCs \citep{2021shenA}, we obtain R$_{GC}$=25.9\arcsec. Adding the GC candidates suggested in \cite{Montes2021} changes this parameter to R$_{GC}$=20.6\arcsec. 
The effective radius of NGC1052-DF2 varies from 22.1\arcsec\ (SDSS; shallower data) to 25.6\arcsec\ (Gemini; deeper data). Consequently, considering all these possible variations, we find that 0.8$<$R$_{GC}$/R$_e$$<$1.2. These low values for the ratio of these two quantities are also found in other ultra diffuse galaxies (UDGs) that show no evidence of tidal distortions \citep{2021MNRAS.502.5921S,2022MNRAS.511.4633S}. In short, both very deep imaging from Gemini plus the GCs distribution do not favor a scenario where the low dark matter content of NGC1052-DF2 can be associated with a removal of the dark matter by tidal forces.

\subsection{Which galaxy is disrupting NGC1052-DF4?}
\label{sub:tidaltailsDF4}

There is a consensus in the community that the excess of light in the outer parts of NGC1052-DF4 found by \citet{Montes2020} can be interpreted as this galaxy being tidally perturbed by a nearby massive neighbor. In support of this scenario, the analysis of the GC distribution of this object also shows that it is significantly extended compared to other galaxies that show no sign of being tidally perturbed (see the right panel of Fig. \ref{fig:globularclusters}). In fact, the ratio of the radius containing half of the confirmed \citep{discoverydf4} and candidate \citep{Montes2020} GCs to the effective radius of this galaxy (R$_{eff,r}$ from Gemini data using fixed ellipses in this work) is R$_{GC}$/R$_e$$\sim$1.7. 
If we assume only spectroscopically confirmed GCs we obtain the same R$_{GC}$ = 41.7\arcsec, which would lead to the same ratio of 1.7.

The above ratio is a lower limit. We can use the effective radius measured in shallower data such as the SDSS (see Sect. \ref{sub:sbdf4}) as a more representative measure of the bulk of the galaxy that is not affected by the light from the object in the tidal tails. When we do this, the ratio between the two radii increases to R$_{GC}$/R$_{eff}$$\sim$2.7. This is around a factor of 2.5 larger than what we observe in UDGs that do not show tidal distortions \citep{2022MNRAS.511.4633S}.

While there is no debate about the excess of light in the outer parts of NGC1052-DF4, there is still no consensus about which galaxy is producing the tidal field responsible for perturbing the system. \citet{Montes2020} propose that the massive galaxy causing the disruption of NGC1052-DF4 is NGC1035 (see Fig. \ref{fig:dithering-df4}). However, \citet{Keim2022} suggest that the disturbance is generated by the massive elliptical NGC1052. There is no way to answer this question definitively without a precise measurement of the distance to all galaxies in the NGC1052 group (not just NGC1052-DF2 and NGC1052-DF4). A comprehensive analysis of the distances to the galaxies in the line of sight of NGC1052 by \citet{monellitrujillo} suggests that there are two groups in the projection. The study suggests that both NGC1052-DF2 and NGC1052-DF4 belong to the foreground group, at 13.5 Mpc, with the more massive galaxies NGC1042 and NGC1035. This would imply that the disruptor is NGC1035. Further evidence for this can be found in \citet{2021roman}. This result is not helped by the latest published distance to NGC1052-DF4, which is D$=$20.0$\pm$1.6 Mpc \citep[][]{2020ApJ...895L...4D}, used by \citet{Keim2022} to assign it to NGC1052. However, to make the scenario even more complex, this distance could be in tension with the recent accurate distance estimate to the most massive galaxy in the group NGC1052 \citep[D=17.9$^{+0.3}_{-0.6}$ Mpc; ][]{2023arXiv230911603J}, which implies a relative distance of $\sim$2.1 Mpc, larger than the virial radius of NGC1052 \citep[390 kpc][]{Forbes2019}. Consequently, the relationship between NGC1052 and NGC1052-DF4, as well as between NGC1035 and NGC1052-DF4, is still not clear. The distance debate is therefore far from settled and will require further work in the future.To help resolve this debate, deep data such as those presented in this paper, but with a larger area coverage that includes NGC1035 in the same pointing, would also be useful to test whether NGC1035 also shows signs of perturbations that could be related to a direct gravitational interaction with NGC1052-DF4.

\section{Conclusion}
\label{sec:conclusions}

The low velocity dispersion of both GCs and stars in NGC1052-DF2 and NGC1052-DF4 \citep{2023shen} may indicate that the dark matter content of these galaxies is low or zero. One possible explanation for this is that the dark matter in these galaxies has been removed via interactions with other objects \citep{ogiya2018, 2020Jackson, maccio2021, ogiya2022}. These mechanisms should leave an excess of brightness in the form of tidal tails in the outer part of the galaxies \citep[e.g.,][]{Johnston2002,2022Moreno}. With sufficiently deep imaging  ($\mu_V$$\sim$30.5 mag/arcsec$^2$), these tidal tails could be detected \cite[see, e.g.,][]{2022Moreno,2023KATAYAMA}. In the current work, we explore this scenario using new ultra-deep images of these galaxies taken with the Gemini telescopes, 1 mag deeper than those previously published.

The new ultra-deep images of NGC1052-DF4 confirm the presence of tidal tails already observed in previous works \citep{Montes2020,Keim2022}. However, the debate continues as to which galaxy is responsible for the tidal destruction of this object. To solve this puzzle it would be necessary to know the exact distance to all the galaxies in the FoV of NGC1052.
In the case of NGC1052-DF2, the new Gemini images show no evidence that the galaxy is being destroyed, to a surface brightness limit of $\mu_g$ = 30.9 mag/arcsec$^2$ (3$\sigma$; 10\arcsec $\times$ 10\arcsec boxes). This observation, together with the fact that the distribution of GCs in this galaxy is  compact (R$_{GC}$/R$_{eff}$$\sim$ 1), suggests that the possible absence of dark matter in this galaxy is not due to its removal. Therefore, either there are no known mechanisms that can sustain the dynamics of this galaxy without dark matter, or the absence of dark matter could be due to an incorrect estimation of the distance to this galaxy \citep{ discoverydf2, vandokkum2018b, 2019Trujillo,monellitrujillo,zonoozi} or an incorrect hypothesis about the dynamical equilibrium of this galaxy \citep{lewis2020,Montes2021}. Further analysis in this regard may help resolve this conflict.

\begin{acknowledgements}

We thank the referee for a careful and constructive reading of the manuscript, which helped to improve the clarity and quality of the presentation.

Based on observations obtained at the international Gemini Observatory, a program of NSF’s NOIRLab, which is managed by the Association of Universities for Research in Astronomy (AURA) under a cooperative agreement with the National Science Foundation on behalf of the Gemini Observatory partnership: the National Science Foundation (United States), National Research Council (Canada), Agencia Nacional de Investigaci\'{o}n y Desarrollo (Chile), Ministerio de Ciencia, Tecnolog\'{i}a e Innovaci\'{o}n (Argentina), Minist\'{e}rio da Ci\^{e}ncia, Tecnologia, Inova\c{c}\~{o}es e Comunica\c{c}\~{o}es (Brazil), and Korea Astronomy and Space Science Institute (Republic of Korea).

The Legacy Surveys consist of three individual and complementary projects: the Dark Energy Camera Legacy Survey (DECaLS; Proposal ID 2014B-0404; PIs: David Schlegel and Arjun Dey), the Beijing-Arizona Sky Survey (BASS; NOAO Prop. ID 2015A-0801; PIs: Zhou Xu and Xiaohui Fan), and the Mayall z-band Legacy Survey (MzLS; Prop. ID 2016A-0453; PI: Arjun Dey). DECaLS, BASS and MzLS together include data obtained, respectively, at the Blanco telescope, Cerro Tololo Inter-American Observatory, NSF’s NOIRLab; the Bok telescope, Steward Observatory, University of Arizona; and the Mayall telescope, Kitt Peak National Observatory, NOIRLab. Pipeline processing and analyses of the data were supported by NOIRLab and the Lawrence Berkeley National Laboratory (LBNL). The Legacy Surveys project is honored to be permitted to conduct astronomical research on Iolkam Du’ag (Kitt Peak), a mountain with particular significance to the Tohono O’odham Nation. NOIRLab is operated by the Association of Universities for Research in Astronomy (AURA) under a cooperative agreement with the National Science Foundation. LBNL is managed by the Regents of the University of California under contract to the U.S. Department of Energy.

\\
MM acknowledges support from the Project PCI2021-122072-2B, financed by MICIN/AEI/10.13039/501100011033, and the European Union “NextGenerationEU”/RTRP.
\\
IT acknowledges support from the ACIISI, Consejer\'{i}a de Econom\'{i}a, Conocimiento y Empleo del Gobierno de Canarias and the European Regional Development Fund (ERDF) under grant with reference PROID2021010044 and from the State Research Agency (AEI-MCINN) of the Spanish Ministry of Science and Innovation under the grant PID2022-140869NB-I00, financed by the Ministry of Science and Innovation, through the State Budget and by the Canary Islands Department of Economy, Knowledge and Employment, through the Regional Budget of the Autonomous Community.
\\
GG acknowledges support from IAC project P/302304 and through the PID2022-140869NB-I00 grant from the Spanish Ministry of Science and Innovation which is partially supported through the state budget and the regional budget of the Consejería de Economía, Industria, Comercio y Conocimiento of the Canary Islands Autonomous Community.
\\
MM and IT acknowledge support from IAC project P/302302.

\textit{Facilities}: 
Gemini: GMOS-S (Program  ID: GS-2021B-FT-104), GMOS-N (Program  ID: GN-2021B-FT-108)

\textit{Software}:
SExtractor          \citep{1996A&AS..117..393B}
SCAMP               \citep{2006ASPC..351..112B}
SWarp               \citep{swarp2010}
\texttt{Gnuastro}            \citep{gnuastro} 
Photutils           \citep{Bradley2019}
Matplotlib          \citep{matplotlib} 
NumPy               \citep{numpy}
Astropy             \citep{Astropy2018}
\end{acknowledgements}

\bibliographystyle{aa}
\bibliography{authors_check} 

\begin{thebibliography}{61}
\expandafter\ifx\csname natexlab\endcsname\relax\def\natexlab#1{#1}\fi

\bibitem[{{Abolfathi} {et~al.}(2018){Abolfathi}, {Aguado}, {Aguilar}, {Allende Prieto}, {Almeida}, {Ananna}, {Anders}, {Anderson}, {Andrews}, {Anguiano}, {Arag{\'o}n-Salamanca}, {Argudo-Fern{\'a}ndez}, {Armengaud}, {Ata}, {Aubourg}, {Avila-Reese}, {Badenes}, {Bailey}, {Balland}, {Barger}, {Barrera-Ballesteros}, {Bartosz}, {Bastien}, {Bates}, {Baumgarten}, {Bautista}, {Beaton}, {Beers}, {Belfiore}, {Bender}, {Bernardi}, {Bershady}, {Beutler}, {Bird}, {Bizyaev}, {Blanc}, {Blanton}, {Blomqvist}, {Bolton}, {Boquien}, {Borissova}, {Bovy}, {Bradna Diaz}, {Brandt}, {Brinkmann}, {Brownstein}, {Bundy}, {Burgasser}, {Burtin}, {Busca}, {Ca{\~n}as}, {Cano-D{\'\i}az}, {Cappellari}, {Carrera}, {Casey}, {Cervantes Sodi}, {Chen}, {Cherinka}, {Chiappini}, {Choi}, {Chojnowski}, {Chuang}, {Chung}, {Clerc}, {Cohen}, {Comerford}, {Comparat}, {Correa do Nascimento}, {da Costa}, {Cousinou}, {Covey}, {Crane}, {Cruz-Gonzalez}, {Cunha}, {da Silva Ilha}, {Damke}, {Darling}, {Davidson}, {Dawson}, {de Icaza Lizaola}, {de la Macorra}, {de
  la Torre}, {De Lee}, {de Sainte Agathe}, {Deconto Machado}, {Dell'Agli}, {Delubac}, {Diamond-Stanic}, {Donor}, {Downes}, {Drory}, {du Mas des Bourboux}, {Duckworth}, {Dwelly}, {Dyer}, {Ebelke}, {Davis Eigenbrot}, {Eisenstein}, {Elsworth}, {Emsellem}, {Eracleous}, {Erfanianfar}, {Escoffier}, {Fan}, {Fern{\'a}ndez Alvar}, {Fernandez-Trincado}, {Fernando Cirolini}, {Feuillet}, {Finoguenov}, {Fleming}, {Font-Ribera}, {Freischlad}, {Frinchaboy}, {Fu}, {G{\'o}mez Maqueo Chew}, {Galbany}, {Garc{\'\i}a P{\'e}rez}, {Garcia-Dias}, {Garc{\'\i}a-Hern{\'a}ndez}, {Garma Oehmichen}, {Gaulme}, {Gelfand}, {Gil-Mar{\'\i}n}, {Gillespie}, {Goddard}, {Gonz{\'a}lez Hern{\'a}ndez}, {Gonzalez-Perez}, {Grabowski}, {Green}, {Grier}, {Gueguen}, {Guo}, {Guy}, {Hagen}, {Hall}, {Harding}, {Hasselquist}, {Hawley}, {Hayes}, {Hearty}, {Hekker}, {Hernandez}, {Hernandez Toledo}, {Hogg}, {Holley-Bockelmann}, {Holtzman}, {Hou}, {Hsieh}, {Hunt}, {Hutchinson}, {Hwang}, {Jimenez Angel}, {Johnson}, {Jones}, {J{\"o}nsson}, {Jullo}, {Khan},
  {Kinemuchi}, {Kirkby}, {Kirkpatrick}, {Kitaura}, {Knapp}, {Kneib}, {Kollmeier}, {Lacerna}, {Lane}, {Lang}, {Law}, {Le Goff}, {Lee}, {Li}, {Li}, {Lian}, {Liang}, {Lima}, {Lin}, {Long}, {Lucatello}, {Lundgren}, {Mackereth}, {MacLeod}, {Mahadevan}, {Maia}, {Majewski}, {Manchado}, {Maraston}, {Mariappan}, {Marques-Chaves}, {Masseron}, {Masters}, {McDermid}, {McGreer}, {Melendez}, {Meneses-Goytia}, {Merloni}, {Merrifield}, {Meszaros}, {Meza}, {Minchev}, {Minniti}, {Mueller}, {Muller-Sanchez}, {Muna}, {Mu{\~n}oz}, {Myers}, {Nair}, {Nandra}, {Ness}, {Newman}, {Nichol}, {Nidever}, {Nitschelm}, {Noterdaeme}, {O'Connell}, {Oelkers}, {Oravetz}, {Oravetz}, {Ort{\'\i}z}, {Osorio}, {Pace}, {Padilla}, {Palanque-Delabrouille}, {Palicio}, {Pan}, {Pan}, {Parikh}, {P{\^a}ris}, {Park}, {Peirani}, {Pellejero-Ibanez}, {Penny}, {Percival}, {Perez-Fournon}, {Petitjean}, {Pieri}, {Pinsonneault}, {Pisani}, {Prada}, {Prakash}, {Queiroz}, {Raddick}, {Raichoor}, {Barboza Rembold}, {Richstein}, {Riffel}, {Riffel}, {Rix}, {Robin},
  {Rodr{\'\i}guez Torres}, {Rom{\'a}n-Z{\'u}{\~n}iga}, {Ross}, {Rossi}, {Ruan}, {Ruggeri}, {Ruiz}, {Salvato}, {S{\'a}nchez}, {S{\'a}nchez}, {Sanchez Almeida}, {S{\'a}nchez-Gallego}, {Santana Rojas}, {Santiago}, {Schiavon}, {Schimoia}, {Schlafly}, {Schlegel}, {Schneider}, {Schuster}, {Schwope}, {Seo}, {Serenelli}, {Shen}, {Shen}, {Shetrone}, {Shull}, {Silva Aguirre}, {Simon}, {Skrutskie}, {Slosar}, {Smethurst}, {Smith}, {Sobeck}, {Somers}, {Souter}, {Souto}, {Spindler}, {Stark}, {Stassun}, {Steinmetz}, {Stello}, {Storchi-Bergmann}, {Streblyanska}, {Stringfellow}, {Su{\'a}rez}, {Sun}, {Szigeti}, {Taghizadeh-Popp}, {Talbot}, {Tang}, {Tao}, {Tayar}, {Tembe}, {Teske}, {Thakar}, {Thomas}, {Tissera}, {Tojeiro}, {Tremonti}, {Troup}, {Urry}, {Valenzuela}, {van den Bosch}, {Vargas-Gonz{\'a}lez}, {Vargas-Maga{\~n}a}, {Vazquez}, {Villanova}, {Vogt}, {Wake}, {Wang}, {Weaver}, {Weijmans}, {Weinberg}, {Westfall}, {Whelan}, {Wilcots}, {Wild}, {Williams}, {Wilson}, {Wood-Vasey}, {Wylezalek}, {Xiao}, {Yan}, {Yang}, {Ybarra},
  {Y{\`e}che}, {Zakamska}, {Zamora}, {Zarrouk}, {Zasowski}, {Zhang}, {Zhao}, {Zhao}, {Zheng}, {Zheng}, {Zhou}, {Zhu}, {Zinn}, \& {Zou}}]{2018sdss}
{Abolfathi}, B., {Aguado}, D.~S., {Aguilar}, G., {et~al.} 2018, \apjs, 235, 42

\bibitem[{{Akhlaghi}(2019)}]{noisechisel_segment_2019}
{Akhlaghi}, M. 2019, arXiv e-prints, arXiv:1909.11230

\bibitem[{{Akhlaghi} \& {Ichikawa}(2015)}]{gnuastro}
{Akhlaghi}, M. \& {Ichikawa}, T. 2015, ApJS, 220, 1

\bibitem[{{Bakos} {et~al.}(2008){Bakos}, {Trujillo}, \& {Pohlen}}]{Bakos2008}
{Bakos}, J., {Trujillo}, I., \& {Pohlen}, M. 2008, \apjl, 683, L103

\bibitem[{{Bennet} {et~al.}(2018){Bennet}, {Sand}, {Zaritsky}, {Crnojevi{\'c}}, {Spekkens}, \& {Karunakaran}}]{2018Bennet}
{Bennet}, P., {Sand}, D.~J., {Zaritsky}, D., {et~al.} 2018, \apjl, 866, L11

\bibitem[{{Bertin}(2006)}]{2006ASPC..351..112B}
{Bertin}, E. 2006, in Astronomical Society of the Pacific Conference Series, Vol. 351, Astronomical Data Analysis Software and Systems XV, ed. C.~{Gabriel}, C.~{Arviset}, D.~{Ponz}, \& S.~{Enrique}, 112

\bibitem[{{Bertin}(2010)}]{swarp2010}
{Bertin}, E. 2010, {SWarp: Resampling and Co-adding FITS Images Together}, Astrophysics Source Code Library, record ascl:1010.068

\bibitem[{{Bertin} \& {Arnouts}(1996)}]{1996A&AS..117..393B}
{Bertin}, E. \& {Arnouts}, S. 1996, \aaps, 117, 393

\bibitem[{Bradley {et~al.}(2019)Bradley, Sipocz, Robitaille, Tollerud, Vinícius, Deil, Barbary, Busko, Günther, Cara, Wilson, Conseil, Droettboom, Bostroem, Bray, Bratholm, Lim, Craig, Barentsen, Pascual, Donath, Greco, Perren, Kerzendorf, de~Val-Borro, Dencheva, de~Albernaz~Ferreira, Souchereau, D'Eugenio, \& Weaver}]{Bradley2019}
Bradley, L., Sipocz, B., Robitaille, T., {et~al.} 2019, astropy/photutils: v0.7.1

\bibitem[{{Busko}(1996)}]{errorsellipse}
{Busko}, I.~C. 1996, in Astronomical Society of the Pacific Conference Series, Vol. 101, Astronomical Data Analysis Software and Systems V, ed. G.~H. {Jacoby} \& J.~{Barnes}, 139

\bibitem[{{Chabrier}(2003)}]{chabrier2003}
{Chabrier}, G. 2003, \pasp, 115, 763

\bibitem[{{Chamba} {et~al.}(2022){Chamba}, {Trujillo}, \& {Knapen}}]{2022nushkia}
{Chamba}, N., {Trujillo}, I., \& {Knapen}, J.~H. 2022, \aap, 667, A87

\bibitem[{{Chowdhury}(2019)}]{2019MNRAS.482L..99C}
{Chowdhury}, A. 2019, \mnras, 482, L99

\bibitem[{{Danieli} {et~al.}(2020){Danieli}, {van Dokkum}, {Abraham}, {Conroy}, {Dolphin}, \& {Romanowsky}}]{2020ApJ...895L...4D}
{Danieli}, S., {van Dokkum}, P., {Abraham}, R., {et~al.} 2020, \apjl, 895, L4

\bibitem[{{Emsellem} {et~al.}(2019){Emsellem}, {van der Burg}, {Fensch}, {Je{\v{r}}{\'a}bkov{\'a}}, {Zanella}, {Agnello}, {Hilker}, {M{\"u}ller}, {Rejkuba}, {Duc}, {Durrell}, {Habas}, {Lelli}, {Lim}, {Marleau}, {Peng}, \& {S{\'a}nchez-Janssen}}]{Emsellem2019}
{Emsellem}, E., {van der Burg}, R. F.~J., {Fensch}, J., {et~al.} 2019, \aap, 625, A76

\bibitem[{{Famaey} {et~al.}(2018){Famaey}, {McGaugh}, \& {Milgrom}}]{2018Famaey}
{Famaey}, B., {McGaugh}, S., \& {Milgrom}, M. 2018, \mnras, 480, 473

\bibitem[{{Fensch} {et~al.}(2019){Fensch}, {van der Burg}, {Je{\v{r}}{\'a}bkov{\'a}}, {Emsellem}, {Zanella}, {Agnello}, {Hilker}, {M{\"u}ller}, {Rejkuba}, {Duc}, {Durrell}, {Habas}, {Lim}, {Marleau}, {Peng}, \& {S{\'a}nchez Janssen}}]{2019fensch}
{Fensch}, J., {van der Burg}, R. F.~J., {Je{\v{r}}{\'a}bkov{\'a}}, T., {et~al.} 2019, \aap, 625, A77

\bibitem[{{Forbes} {et~al.}(2019){Forbes}, {Alabi}, {Brodie}, \& {Romanowsky}}]{Forbes2019}
{Forbes}, D.~A., {Alabi}, A., {Brodie}, J.~P., \& {Romanowsky}, A.~J. 2019, \mnras, 489, 3665

\bibitem[{{Hook} {et~al.}(2004){Hook}, {J{\o}rgensen}, {Allington-Smith}, {Davies}, {Metcalfe}, {Murowinski}, \& {Crampton}}]{2004hook}
{Hook}, I.~M., {J{\o}rgensen}, I., {Allington-Smith}, J.~R., {et~al.} 2004, \pasp, 116, 425

\bibitem[{{Hunter}(2007)}]{matplotlib}
{Hunter}, J.~D. 2007, Computing in Science and Engineering, 9, 90

\bibitem[{{Infante-Sainz} \& {Akhlaghi}(2024)}]{scriptcolorimagesRaul}
{Infante-Sainz}, R. \& {Akhlaghi}, M. 2024, Research Notes of the American Astronomical Society, 8, 10

\bibitem[{{Infante-Sainz} {et~al.}(2024){Infante-Sainz}, {Akhlaghi}, \& {Eskandarlou}}]{astscriptradialprofileRaul}
{Infante-Sainz}, R., {Akhlaghi}, M., \& {Eskandarlou}, S. 2024, Research Notes of the American Astronomical Society, 8, 22

\bibitem[{{Jackson} {et~al.}(2020){Jackson}, {Martin}, {Kaviraj}, {Laigle}, {Devriendt}, {Dubois}, \& {Pichon}}]{2020Jackson}
{Jackson}, R.~A., {Martin}, G., {Kaviraj}, S., {et~al.} 2020, \mnras, 494, 5568

\bibitem[{{Jacoby} {et~al.}(2023){Jacoby}, {Ciardullo}, {Roth}, {Arnaboldi}, \& {Weilbacher}}]{2023arXiv230911603J}
{Jacoby}, G.~H., {Ciardullo}, R., {Roth}, M.~M., {Arnaboldi}, M., \& {Weilbacher}, P.~M. 2023, arXiv e-prints, arXiv:2309.11603

\bibitem[{{Jedrzejewski}(1987)}]{Jedrzejewski1987}
{Jedrzejewski}, R.~I. 1987, \mnras, 226, 747

\bibitem[{{Johnston} {et~al.}(2002){Johnston}, {Choi}, \& {Guhathakurta}}]{Johnston2002}
{Johnston}, K.~V., {Choi}, P.~I., \& {Guhathakurta}, P. 2002, \aj, 124, 127

\bibitem[{{Karachentsev} {et~al.}(2000){Karachentsev}, {Karachentseva}, {Suchkov}, \& {Grebel}}]{2000A&AS..145..415K}
{Karachentsev}, I.~D., {Karachentseva}, V.~E., {Suchkov}, A.~A., \& {Grebel}, E.~K. 2000, \aaps, 145, 415

\bibitem[{{Katayama} \& {Nagamine}(2023)}]{2023KATAYAMA}
{Katayama}, R. \& {Nagamine}, K. 2023, arXiv e-prints, arXiv:2306.07756

\bibitem[{{Keim} {et~al.}(2022){Keim}, {Dokkum}, {Danieli}, {Lokhorst}, {Li}, {Shen}, {Abraham}, {Chen}, {Gilhuly}, {Liu}, {Merritt}, {Miller}, {Pasha}, \& {Polzin}}]{Keim2022}
{Keim}, M.~A., {Dokkum}, P.~v., {Danieli}, S., {et~al.} 2022, \apj, 935, 160

\bibitem[{{Kroupa} {et~al.}(2018){Kroupa}, {Haghi}, {Javanmardi}, {Zonoozi}, {M{\"u}ller}, {Banik}, {Wu}, {Zhao}, \& {Dabringhausen}}]{2018Kroupamond}
{Kroupa}, P., {Haghi}, H., {Javanmardi}, B., {et~al.} 2018, \nat, 561, E4

\bibitem[{{Lewis} {et~al.}(2020){Lewis}, {Brewer}, \& {Wan}}]{lewis2020}
{Lewis}, G.~F., {Brewer}, B.~J., \& {Wan}, Z. 2020, \mnras, 491, L1

\bibitem[{{Macci{\`o}} {et~al.}(2021){Macci{\`o}}, {Prats}, {Dixon}, {Buck}, {Waterval}, {Arora}, {Courteau}, \& {Kang}}]{maccio2021}
{Macci{\`o}}, A.~V., {Prats}, D.~H., {Dixon}, K.~L., {et~al.} 2021, \mnras, 501, 693

\bibitem[{{Martin} {et~al.}(2018){Martin}, {Collins}, {Longeard}, \& {Tollerud}}]{2018Martin}
{Martin}, N.~F., {Collins}, M. L.~M., {Longeard}, N., \& {Tollerud}, E. 2018, \apjl, 859, L5

\bibitem[{{Monelli} \& {Trujillo}(2019)}]{monellitrujillo}
{Monelli}, M. \& {Trujillo}, I. 2019, \apjl, 880, L11

\bibitem[{{Montes} {et~al.}(2020){Montes}, {Infante-Sainz}, {Madrigal-Aguado}, {Rom{\'a}n}, {Monelli}, {Borlaff}, \& {Trujillo}}]{Montes2020}
{Montes}, M., {Infante-Sainz}, R., {Madrigal-Aguado}, A., {et~al.} 2020, \apj, 904, 114

\bibitem[{{Montes} {et~al.}(2021){Montes}, {Trujillo}, {Infante-Sainz}, {Monelli}, \& {Borlaff}}]{Montes2021}
{Montes}, M., {Trujillo}, I., {Infante-Sainz}, R., {Monelli}, M., \& {Borlaff}, A.~S. 2021, \apj, 919, 56

\bibitem[{{Moreno} {et~al.}(2022){Moreno}, {Danieli}, {Bullock}, {Feldmann}, {Hopkins}, {{\c{c}}atmabacak}, {Gurvich}, {Lazar}, {Klein}, {Hummels}, {Hafen}, {Mercado}, {Yu}, {Jiang}, {Wheeler}, {Wetzel}, {Angl{\'e}s-Alc{\'a}zar}, {Boylan-Kolchin}, {Quataert}, {Faucher-Gigu{\`e}re}, \& {Kere{\v{s}}}}]{2022Moreno}
{Moreno}, J., {Danieli}, S., {Bullock}, J.~S., {et~al.} 2022, Nature Astronomy, 6, 496

\bibitem[{{M{\"u}ller} {et~al.}(2019){M{\"u}ller}, {Rich}, {Rom{\'a}n}, {Y{\i}ld{\i}z}, {B{\'\i}lek}, {Duc}, {Fensch}, {Trujillo}, \& {Koch}}]{muller2019}
{M{\"u}ller}, O., {Rich}, R.~M., {Rom{\'a}n}, J., {et~al.} 2019, \aap, 624, L6

\bibitem[{{Ogiya}(2018)}]{ogiya2018}
{Ogiya}, G. 2018, \mnras, 480, L106

\bibitem[{{Ogiya} {et~al.}(2022){Ogiya}, {van den Bosch}, \& {Burkert}}]{ogiya2022}
{Ogiya}, G., {van den Bosch}, F.~C., \& {Burkert}, A. 2022, \mnras, 510, 2724

\bibitem[{{Roediger} \& {Courteau}(2015)}]{2015MNRAS.452.3209R}
{Roediger}, J.~C. \& {Courteau}, S. 2015, \mnras, 452, 3209

\bibitem[{{Rom{\'a}n} {et~al.}(2021){Rom{\'a}n}, {Castilla}, \& {Pascual-Granado}}]{2021roman}
{Rom{\'a}n}, J., {Castilla}, A., \& {Pascual-Granado}, J. 2021, \aap, 656, A44

\bibitem[{{Rom{\'a}n} {et~al.}(2020){Rom{\'a}n}, {Trujillo}, \& {Montes}}]{Javicirri2020}
{Rom{\'a}n}, J., {Trujillo}, I., \& {Montes}, M. 2020, \aap, 644, A42

\bibitem[{{Ruiz-Lara} {et~al.}(2019){Ruiz-Lara}, {Trujillo}, {Beasley}, {Falc{\'o}n-Barroso}, {Vazdekis}, {Filho}, {Monelli}, {Rom{\'a}n}, \& {S{\'a}nchez Almeida}}]{2019MNRAS.486.5670R}
{Ruiz-Lara}, T., {Trujillo}, I., {Beasley}, M.~A., {et~al.} 2019, \mnras, 486, 5670

\bibitem[{{Saifollahi} {et~al.}(2021){Saifollahi}, {Trujillo}, {Beasley}, {Peletier}, \& {Knapen}}]{2021MNRAS.502.5921S}
{Saifollahi}, T., {Trujillo}, I., {Beasley}, M.~A., {Peletier}, R.~F., \& {Knapen}, J.~H. 2021, \mnras, 502, 5921

\bibitem[{{Saifollahi} {et~al.}(2022){Saifollahi}, {Zaritsky}, {Trujillo}, {Peletier}, {Knapen}, {Amorisco}, {Beasley}, \& {Donnerstein}}]{2022MNRAS.511.4633S}
{Saifollahi}, T., {Zaritsky}, D., {Trujillo}, I., {et~al.} 2022, \mnras, 511, 4633

\bibitem[{{Schirmer}(2013)}]{2013THELI}
{Schirmer}, M. 2013, \apjs, 209, 21

\bibitem[{{Schlafly} \& {Finkbeiner}(2011)}]{2011Schlafly}
{Schlafly}, E.~F. \& {Finkbeiner}, D.~P. 2011, \apj, 737, 103

\bibitem[{{Shen} {et~al.}(2021){Shen}, {van Dokkum}, \& {Danieli}}]{2021shenA}
{Shen}, Z., {van Dokkum}, P., \& {Danieli}, S. 2021, \apj, 909, 179

\bibitem[{{Shen} {et~al.}(2023){Shen}, {van Dokkum}, \& {Danieli}}]{2023shen}
{Shen}, Z., {van Dokkum}, P., \& {Danieli}, S. 2023, arXiv e-prints, arXiv:2309.08592

\bibitem[{{The Astropy Collaboration} {et~al.}(2018){The Astropy Collaboration}, {Price-Whelan}, {Sip{\H o}cz}, {G{\"u}nther}, {Lim}, {Crawford}, {Conseil}, {Shupe}, {Craig}, {Dencheva}, {Ginsburg}, {VanderPlas}, {Bradley}, {P{\'e}rez-Su{\'a}rez}, {de Val-Borro}, {Aldcroft}, {Cruz}, {Robitaille}, {Tollerud}, {Ardelean}, {Babej}, {Bachetti}, {Bakanov}, {Bamford}, {Barentsen}, {Barmby}, {Baumbach}, {Berry}, {Biscani}, {Boquien}, {Bostroem}, {Bouma}, {Brammer}, {Bray}, {Breytenbach}, {Buddelmeijer}, {Burke}, {Calderone}, {Cano Rodr{\'{\i}}guez}, {Cara}, {Cardoso}, {Cheedella}, {Copin}, {Crichton}, {D{\'A}vella}, {Deil}, {Depagne}, {Dietrich}, {Donath}, {Droettboom}, {Earl}, {Erben}, {Fabbro}, {Ferreira}, {Finethy}, {Fox}, {Garrison}, {Gibbons}, {Goldstein}, {Gommers}, {Greco}, {Greenfield}, {Groener}, {Grollier}, {Hagen}, {Hirst}, {Homeier}, {Horton}, {Hosseinzadeh}, {Hu}, {Hunkeler}, {Ivezi{\'c}}, {Jain}, {Jenness}, {Kanarek}, {Kendrew}, {Kern}, {Kerzendorf}, {Khvalko}, {King}, {Kirkby}, {Kulkarni}, {Kumar},
  {Lee}, {Lenz}, {Littlefair}, {Ma}, {Macleod}, {Mastropietro}, {McCully}, {Montagnac}, {Morris}, {Mueller}, {Mumford}, {Muna}, {Murphy}, {Nelson}, {Nguyen}, {Ninan}, {N{\"o}the}, {Ogaz}, {Oh}, {Parejko}, {Parley}, {Pascual}, {Patil}, {Patil}, {Plunkett}, {Prochaska}, {Rastogi}, {Reddy Janga}, {Sabater}, {Sakurikar}, {Seifert}, {Sherbert}, {Sherwood-Taylor}, {Shih}, {Sick}, {Silbiger}, {Singanamalla}, {Singer}, {Sladen}, {Sooley}, {Sornarajah}, {Streicher}, {Teuben}, {Thomas}, {Tremblay}, {Turner}, {Terr{\'o}n}, {van Kerkwijk}, {de la Vega}, {Watkins}, {Weaver}, {Whitmore}, {Woillez}, \& {Zabalza}}]{Astropy2018}
{The Astropy Collaboration}, {Price-Whelan}, A.~M., {Sip{\H o}cz}, B.~M., {et~al.} 2018, ArXiv e-prints [\eprint[arXiv]{1801.02634}]

\bibitem[{{Trujillo} {et~al.}(2019){Trujillo}, {Beasley}, {Borlaff}, {Carrasco}, {Di Cintio}, {Filho}, {Monelli}, {Montes}, {Rom{\'a}n}, {Ruiz-Lara}, {S{\'a}nchez Almeida}, {Valls-Gabaud}, \& {Vazdekis}}]{2019Trujillo}
{Trujillo}, I., {Beasley}, M.~A., {Borlaff}, A., {et~al.} 2019, \mnras, 486, 1192

\bibitem[{{Trujillo} {et~al.}(2021){Trujillo}, {D'Onofrio}, {Zaritsky}, {Madrigal-Aguado}, {Chamba}, {Golini}, {Akhlaghi}, {Sharbaf}, {Infante-Sainz}, {Rom{\'a}n}, {Morales-Socorro}, {Sand}, \& {Martin}}]{LIGHTSs}
{Trujillo}, I., {D'Onofrio}, M., {Zaritsky}, D., {et~al.} 2021, \aap, 654, A40

\bibitem[{{Trujillo} \& {Fliri}(2016)}]{trujillofliri2016}
{Trujillo}, I. \& {Fliri}, J. 2016, \apj, 823, 123

\bibitem[{{van der Walt} {et~al.}(2011){van der Walt}, {Colbert}, \& {Varoquaux}}]{numpy}
{van der Walt}, S., {Colbert}, S.~C., \& {Varoquaux}, G. 2011, Computing in Science and Engineering, 13, 22

\bibitem[{{van Dokkum} {et~al.}(2019){van Dokkum}, {Danieli}, {Abraham}, {Conroy}, \& {Romanowsky}}]{discoverydf4}
{van Dokkum}, P., {Danieli}, S., {Abraham}, R., {Conroy}, C., \& {Romanowsky}, A.~J. 2019, \apjl, 874, L5

\bibitem[{{van Dokkum} {et~al.}(2018{\natexlab{a}}){van Dokkum}, {Danieli}, {Cohen}, {Merritt}, {Romanowsky}, {Abraham}, {Brodie}, {Conroy}, {Lokhorst}, {Mowla}, {O'Sullivan}, \& {Zhang}}]{discoverydf2}
{van Dokkum}, P., {Danieli}, S., {Cohen}, Y., {et~al.} 2018{\natexlab{a}}, \nat, 555, 629

\bibitem[{{van Dokkum} {et~al.}(2018{\natexlab{b}}){van Dokkum}, {Danieli}, {Cohen}, {Romanowsky}, \& {Conroy}}]{vandokkum2018b}
{van Dokkum}, P., {Danieli}, S., {Cohen}, Y., {Romanowsky}, A.~J., \& {Conroy}, C. 2018{\natexlab{b}}, \apjl, 864, L18

\bibitem[{{van Dokkum} {et~al.}(2022){van Dokkum}, {Shen}, {Keim}, {Trujillo-Gomez}, {Danieli}, {Dutta Chowdhury}, {Abraham}, {Conroy}, {Kruijssen}, {Nagai}, \& {Romanowsky}}]{2022vandokkum}
{van Dokkum}, P., {Shen}, Z., {Keim}, M.~A., {et~al.} 2022, \nat, 605, 435

\bibitem[{{Vazdekis} {et~al.}(2016){Vazdekis}, {Koleva}, {Ricciardelli}, {R{\"o}ck}, \& {Falc{\'o}n-Barroso}}]{vazdekis}
{Vazdekis}, A., {Koleva}, M., {Ricciardelli}, E., {R{\"o}ck}, B., \& {Falc{\'o}n-Barroso}, J. 2016, \mnras, 463, 3409

\bibitem[{{Zonoozi} {et~al.}(2021){Zonoozi}, {Haghi}, \& {Kroupa}}]{zonoozi}
{Zonoozi}, A.~H., {Haghi}, H., \& {Kroupa}, P. 2021, \mnras, 504, 1668

\end{thebibliography}


\begin{appendix}

\section{Dithering pattern and airmass conditions}
\label{app:dithering}

We developed a special observing strategy to deal with vignetting caused by the arm of the guide star and to minimize scattered light effects on GMOS cameras. This strategy involves implementing a dithering pattern with an offset of 10$\arcsec$, which is applied both over the galaxy and over an adjacent region of the sky free of bright objects. For every six images of the galaxy, we repositioned the telescope to take five images of a nearby region of the sky. We then returned the telescope to the original location of the galaxy. This approach allowed us to build the master flat field, a critical component of deep imaging, by combining only these sky images. By adopting this observing strategy, we aim to mitigate vignetting problems, reduce scattered light contamination, and optimize the quality of our imaging data obtained with the Gemini telescopes.

\begin{figure*}
\centering
    \includegraphics[width=0.9\textwidth]{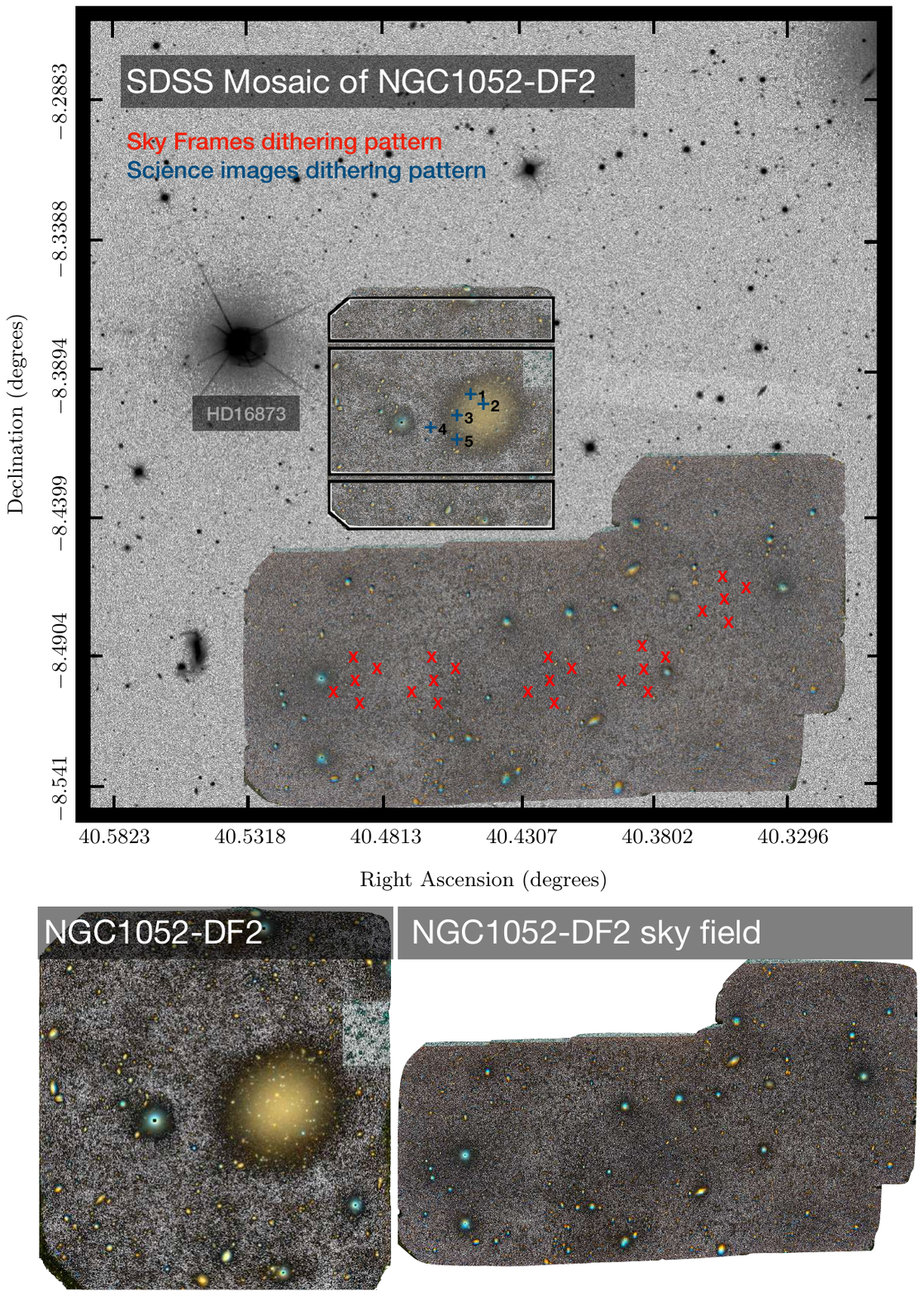}
    \caption{NGC1052-DF2 observational strategy. Top panel: NGC1052-DF2 field created from the SDSS r-band image. The image covers an area of 18$^\prime$$\times$18$^\prime$. The five blue crosses indicate the representative centers of the dithering pattern surrounding the galaxy, while red crosses highlight the central positions of the sky frames. The dimensions of the GMOS CCDs are outlined in black around the galaxy, representing a total size for the camera of 5$^\prime$$\times$5$^\prime$. The name of the bright star (m$_V$ = 8.34 mag) near NGC1052-DF2 is indicated by a label. Lower-left panel: Color composite image (using g and r filters) of NGC1052-DF2 obtained in this work with the Gemini telescope. Lower-right panel: Stack of sky images (g and r filters).}
    \label{fig:dithering-df2}
\end{figure*}

\begin{figure*} 
\centering
    \includegraphics[width=0.9\textwidth]{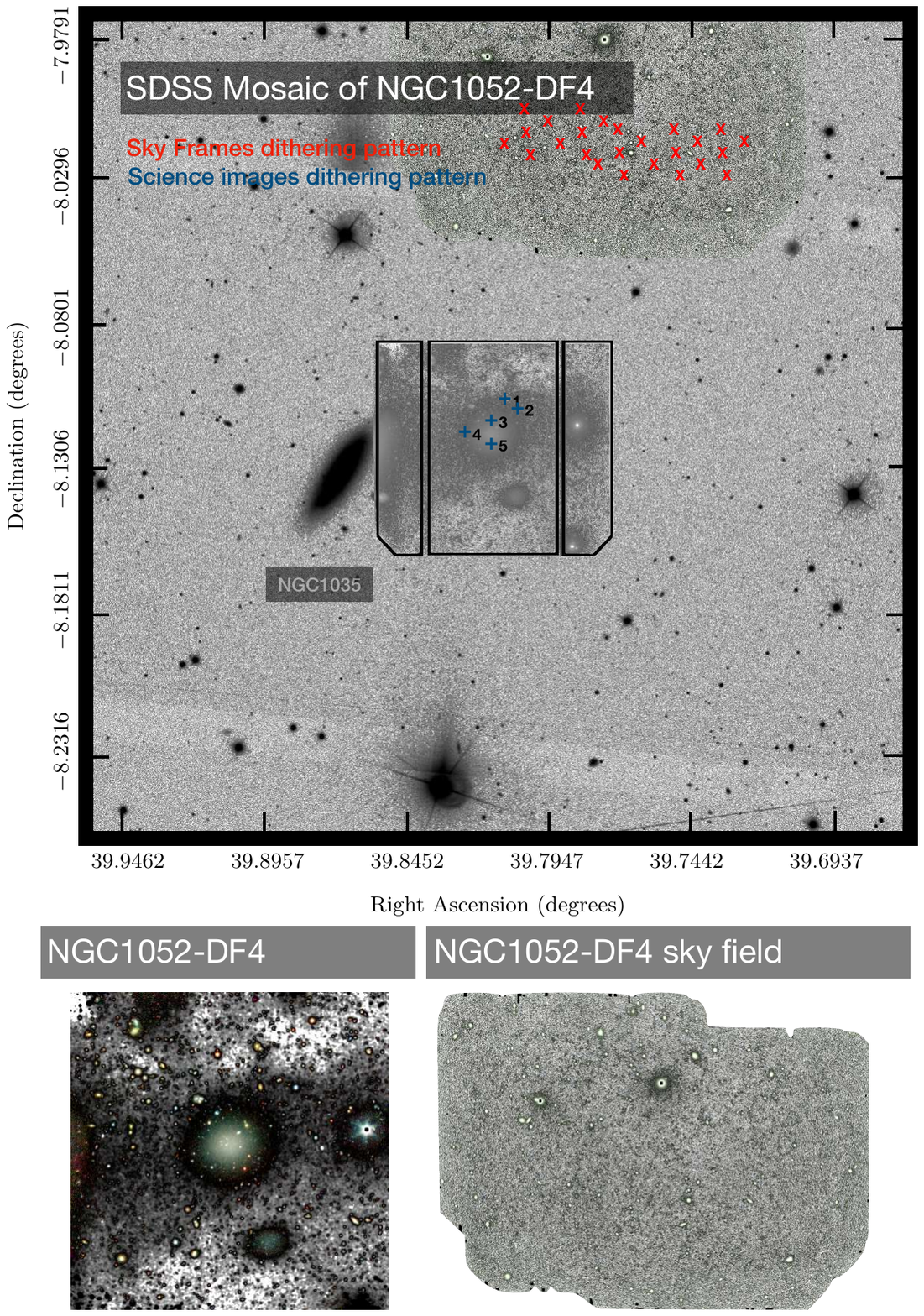}
    \caption{Similar to Fig. \ref{fig:dithering-df2}, but for NGC1052-DF4. Upper panel: NGC1052-DF4 field over an area of 18$^\prime$$\times$18$^\prime$ using data from the SDSS r band. The dimension of the GMOS camera and the centers of the dithering pattern around the galaxy and for the sky images are highlighted. 
    Lower-left panel: Final stack of the galaxy in black and white background, using the GMOS r filter. The color part of the Gemini stamp is created using rescaled HiPERCAM data from \citet{Montes2020} using the g, r, and z filters. Lower-right panel: Sky region that was observed to obtain the flat field of the data.}
    \label{fig:dithering-df4}
\end{figure*}

In the upper panel of Fig. \ref{fig:dithering-df2} we present the SDSS mosaic of the sky region surrounding NGC1052-DF2 (18$^\prime$$\times$18$^\prime$). The background image is created using the SDSS r filter downloaded from the DR12 Science Archive Server (SAS). The CCDs of the Gemini camera (total FoV = 5\arcmin $\times$5\arcmin) are outlined in black on top of the galaxy. Five representative central positions of the dithering pattern over the galaxy are marked with blue crosses. The central coordinates of each sky frame are highlighted in red. The lower-left panel shows the final Gemini stack of galaxy frames, while the lower-right panel shows the final distribution of sky regions used to construct the flat field of the images.  In the same way, Fig. \ref{fig:dithering-df4} illustrates the observing strategy of NGC1052-DF4.


\begin{figure*}
    \centering
    \includegraphics[width=\textwidth]{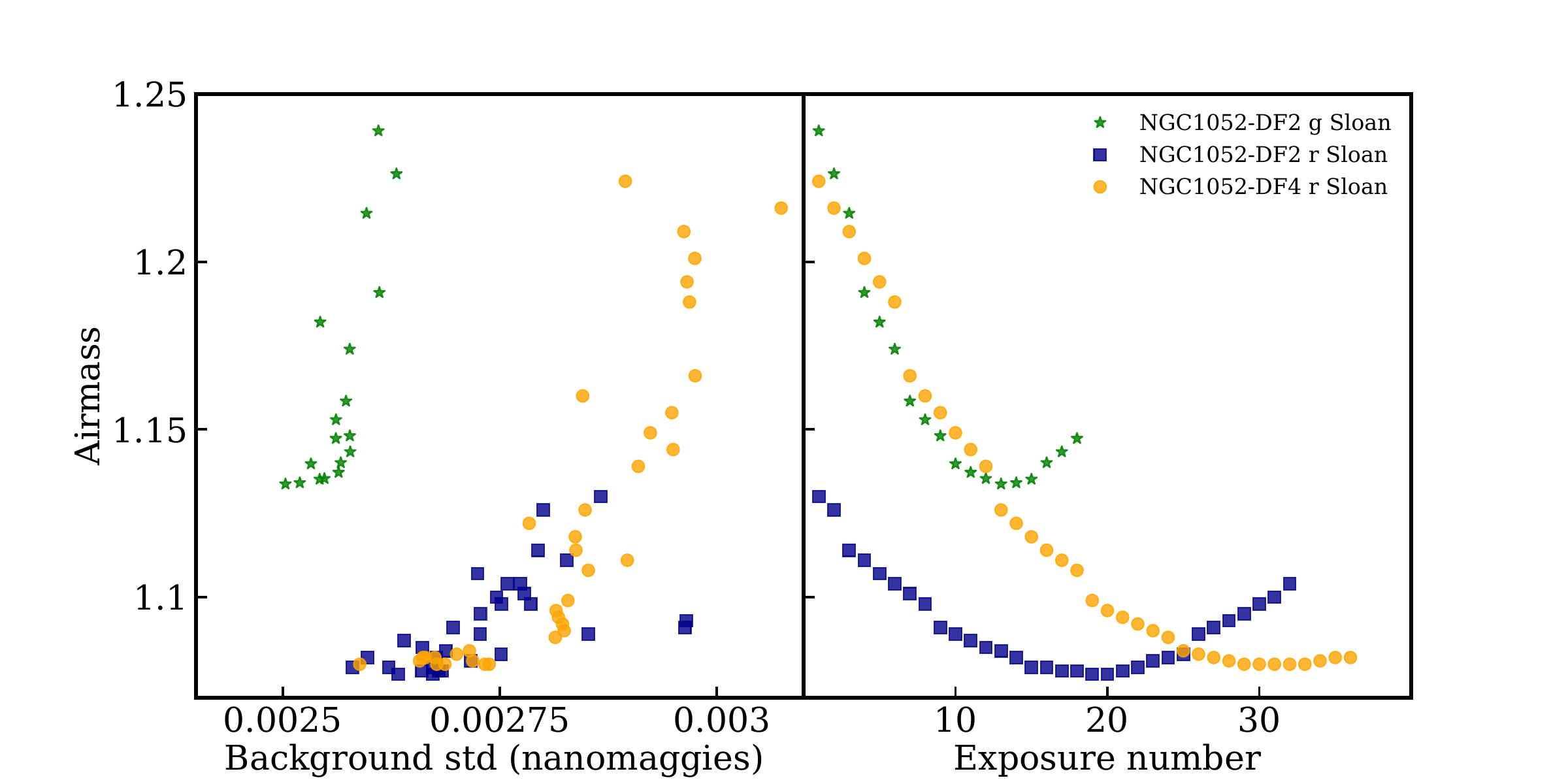}
    \caption{Airmass and sky variation during observations. Left panel: Relationship between image quality (characterized by the standard deviation of the background pixels in nanomaggies units) and the airmass of the observations. Poorer quality images (larger values of the background standard deviation) are observed in those images with larger airmass values. Right panel: Variation in the airmass during the observation run.
}
    \label{ima:airmass}
\end{figure*}

Variations in airmass correspond to differences in image quality, which we can characterize by the standard deviation of the background pixels (in units after photometric calibration, i.e., nanomaggies in our case). The left plot in Fig. \ref{ima:airmass} shows the relationship between the airmass of each science image and the standard deviation of the background of those images. There is a clear trend that larger airmasses are associated with lower-quality images (i.e., larger values of the standard deviation of the background). The right panel of the figure shows the variation of the airmass during the run.

\section{Surface brightness, ellipticity, and position angle profiles}
\label{app:profiles}

\begin{figure*}
    \centering
    \includegraphics[width=\textwidth]{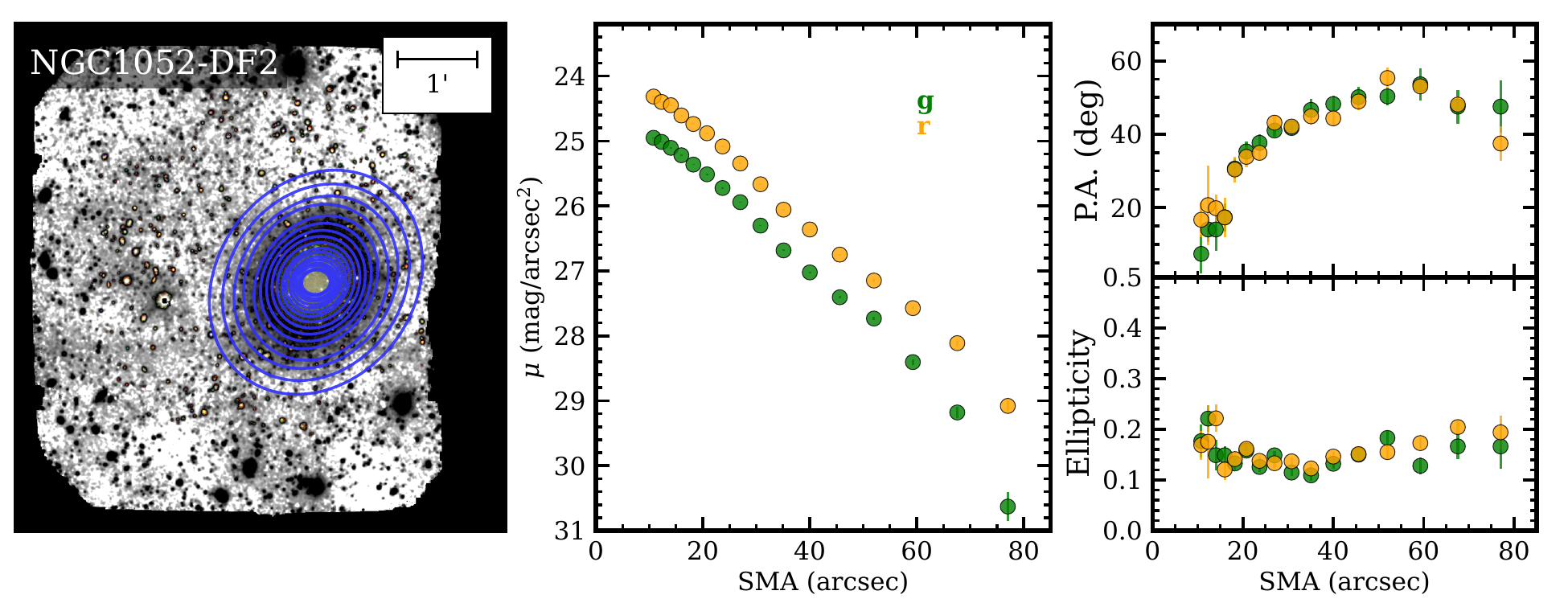}
    \caption{Surface brightness, ellipticity, and PA radial profiles of NGC1052-DF2 obtained with the \texttt{ellipse} routine. The profiles are obtained for both the g and r bands. Left panel: Shape of the ellipses used to compute the profiles over a grayscale background image of g- + r- Sloan Gemini data. The color part of the stamp is created using rescaled HiPERCAM data from \citet{Montes2020} using the g, r, and z filters. Middle panel: g-band (green) and r-band (orange) surface brightness profiles. Errors associated with the surface brightness profiles are computed as explained in Sect. \ref{subsub:sbdf2}.
    Right panel: Radial profile of the PA and ellipticity for the two filters separately down to 80\arcsec. The errors on those parameters are provided by the \texttt{ellipse} routine (see more details in \citet{errorsellipse}).
    }
    \label{fig:outputdf2}
\end{figure*}

Fig. \ref{fig:outputdf2} shows the output of the \texttt{ellipse} routine used to extract the surface brightness profiles of NGC1052-DF2 in the calibrated Sloan g and r bands separately. The figure includes information on the change in ellipticity and PA of the isophotes with radius. The same, but for the galaxy NGC1052-DF4, is shown in Fig. \ref{ima:outputdf4}.

\begin{figure*}
    \centering
    \includegraphics[width=0.5\textwidth]{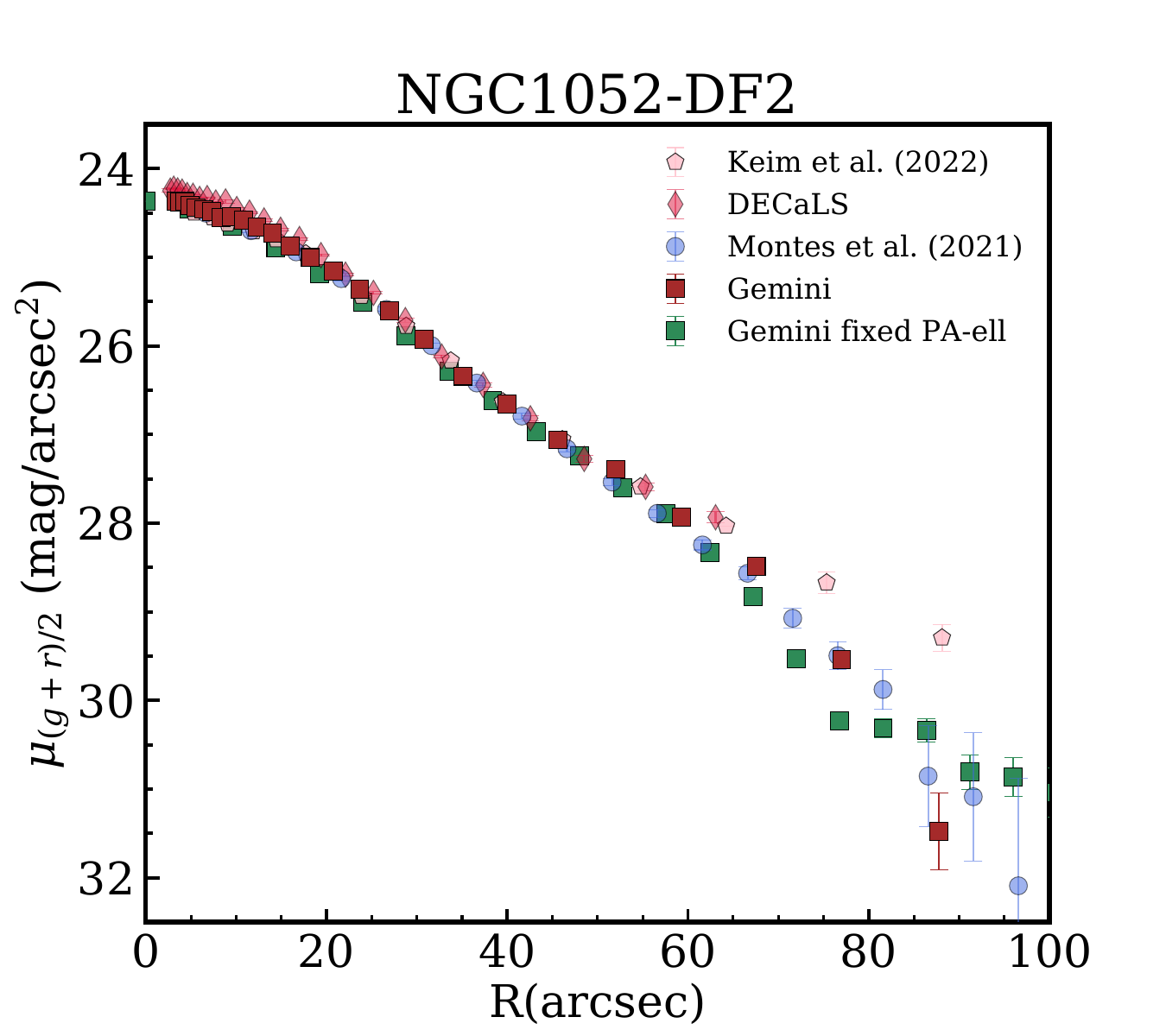}
    \caption{Surface brightness profile of NGC1052-DF2 using different telescopes: INT \citep{Montes2021}, DECaLS (this work), Dragonfly \citep{Keim2022}, and Gemini (this work).
    \citet{Montes2021} derive the profiles with a fixed ellipticity and PA. In the case of Gemini (dark red points), DECaLS, and Dragonfly data, the PA and ellipticity vary freely.
    We also show with green data points the surface brightness profile where ellipticity and PA are fixed to values representative of the outer parts of the galaxy (see the main text for details) for Gemini data.}
    \label{fig:sb-comparison}
\end{figure*}

\begin{figure*}
    \centering
    \includegraphics[width=\textwidth]{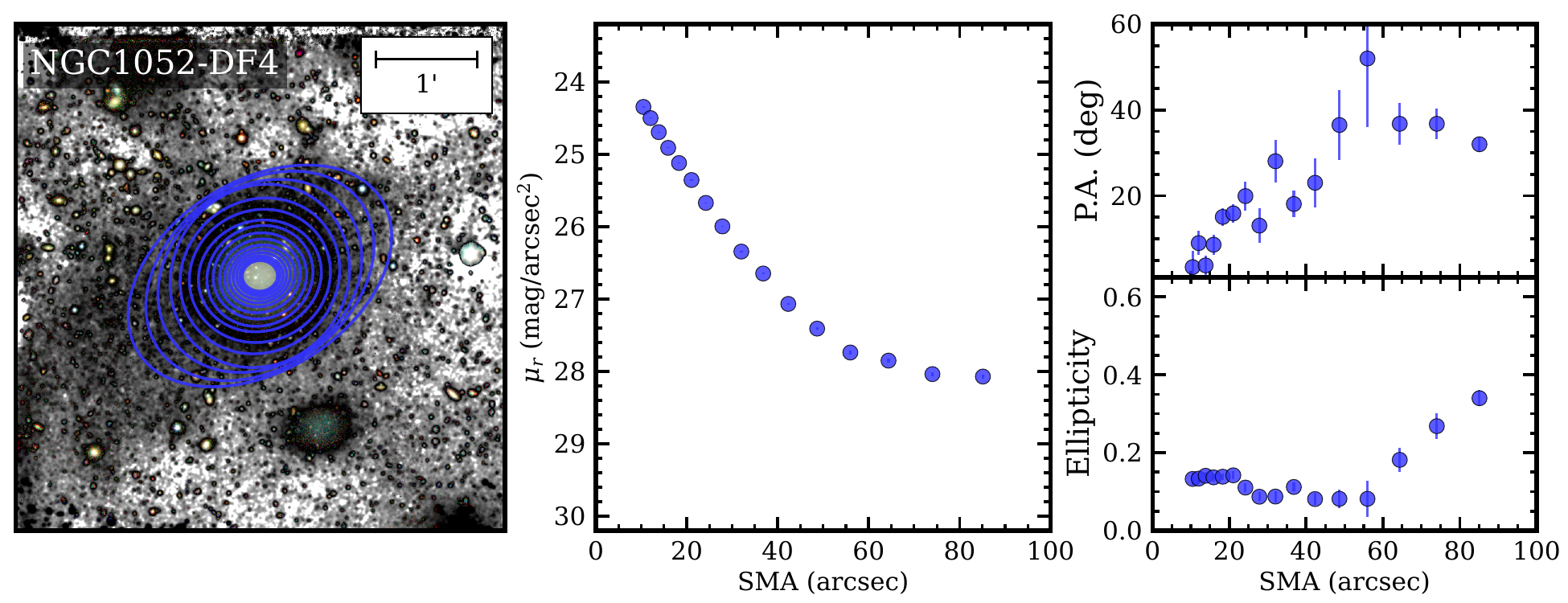}
    \caption{Same as Fig. \ref{fig:outputdf2} but for NGC1052-DF4. Only the r-filter image is available for this galaxy.}
    \label{ima:outputdf4}
\end{figure*}

The effect of obtaining surface brightness profiles of NGC1052-DF2 with varying ellipticity and PA with radius is shown in Fig. \ref{fig:sb-comparison}. The Gemini data (dark red dots) are compared with surface brightness profiles from other telescopes. The profiles correspond to those derived using the g and r images combined to allow comparison with more datasets from the literature. In addition, we show the surface brightness profile for the Gemini data when the ellipticity and PA are kept fixed (green data points). As expected, allowing the ellipticity and PA to vary smooths out the truncation observed when the ellipticity and PA are fixed. Nevertheless, the change in slope is still visible in both the Gemini and INT data. The Dragonfly profile (pink dots) shows no truncation at this radial distance. Considering that the ellipticity and PA derived with such a dataset are very similar to those found with the other telescopes, we think that the most likely explanation for this different behavior is related to the estimation of the background around the galaxy between the different works.

\begin{figure*}
    \centering
    \includegraphics[width=0.5\textwidth]{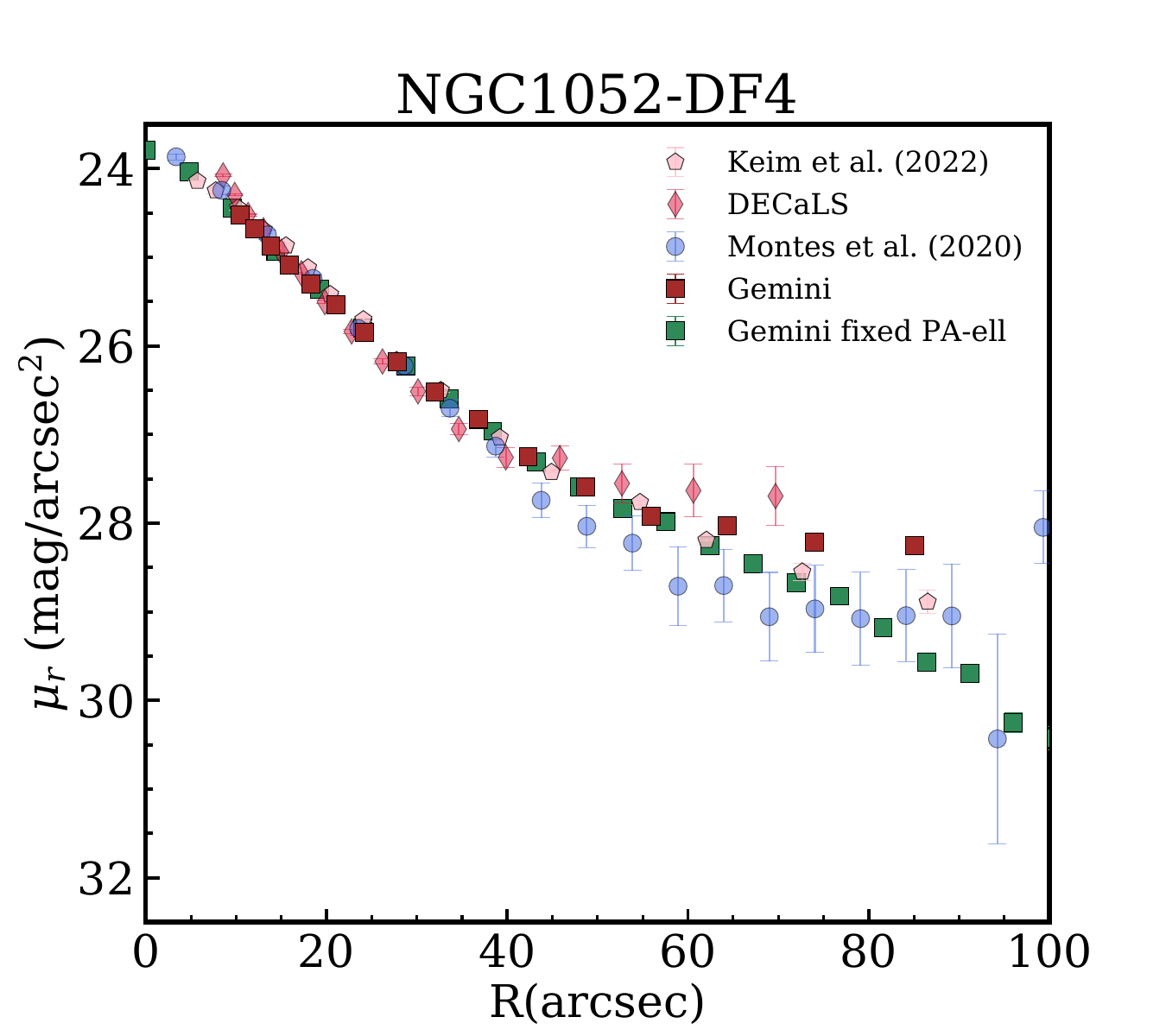}
    \caption{Surface brightness profile of NGC1052-DF4 with different telescopes: IAC80 \citep{Montes2020}, DECaLS (this work), Dragonfly \citep{Keim2022}, and Gemini (this work). In all cases the ellipticity and PA are allowed to vary freely. We also include the Gemini profile (green dots) resulting from fixing the ellipticity and PA (see the main text for details). The IAC80 data correspond to images where the g-, r-, and i-band filters have been combined, and for Dragonfly the g and r bands have been combined, while in the case of DECaLS and Gemini we show only the r-band results. The IAC80 and Dragonfly profiles have been shifted vertically by 0.1 mag and 0.25 mag, respectively, to match the r-band profiles only, to facilitate comparison between different telescopes.}
    \label{ima:sb-comparison-df4}
\end{figure*}

In Fig. \ref{ima:sb-comparison-df4}, we compare the surface brightness profile of NGC1052-DF4 when the ellipses describing the isophotes are left free in axis ratio and PA. On doing this we can compare with previously published results using IAC80 \citep{Montes2020} and Dragonfly \citep{Keim2022}. Despite having different PA in the outer parts, the profiles are remarkably similar. All the works agree on an excess of light in the outer parts of NGC1052-DF4 with a similar slope and at a similar surface brightness level.
 
\section{Masks used in this work}
\label{app:masks} 

\begin{figure*}
    \centering
    \includegraphics[width=\textwidth]{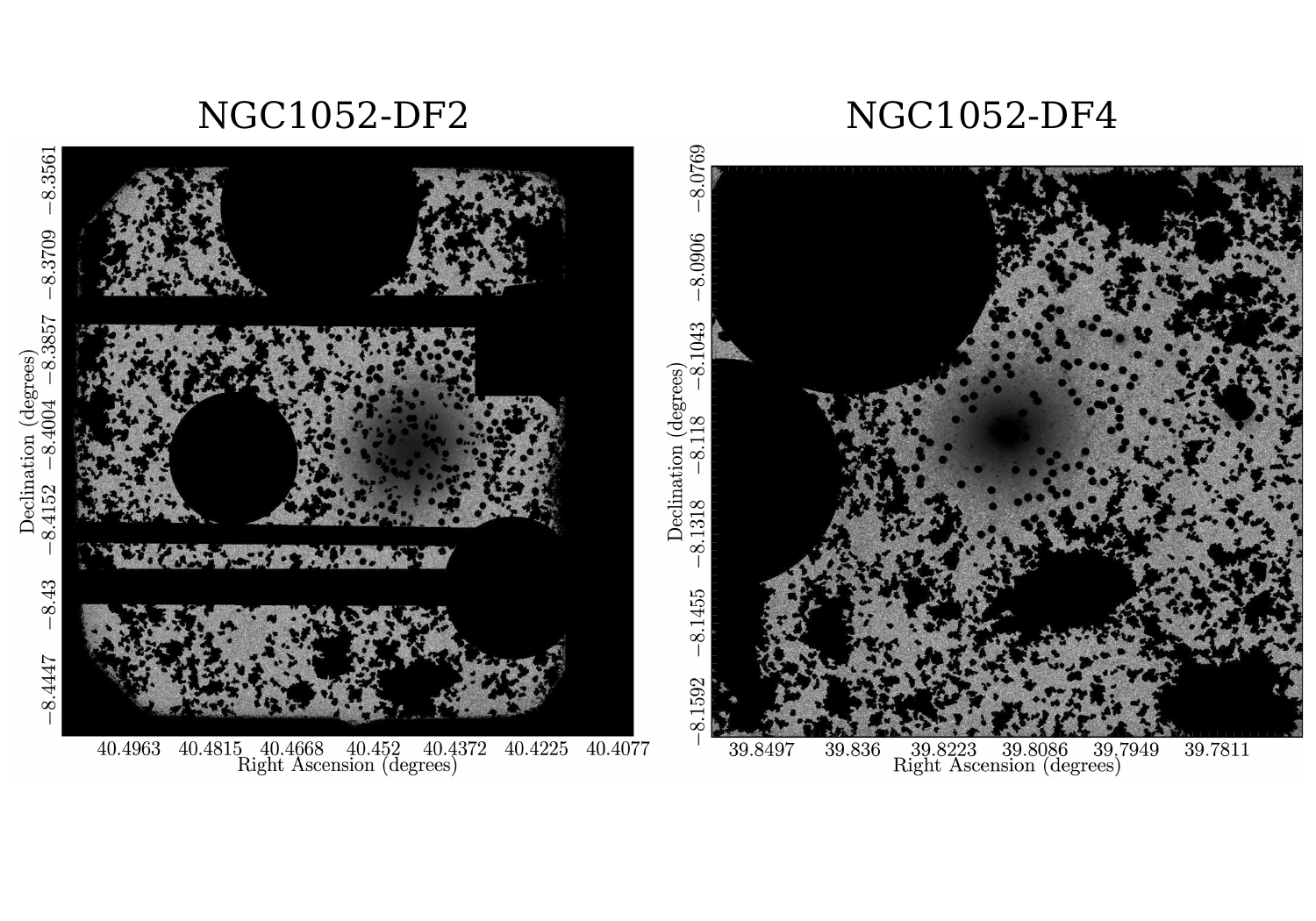}
    \caption{Masks used to compute the surface brightness profiles for NGC1052-DF2 (left) and NGC1052-DF4 (right).}
    \label{ima:mask-df2df4}
\end{figure*}

Figure \ref{ima:mask-df2df4} shows the masks used in both Gemini images to compute the surface brightness profiles.

\end{appendix}

\end{document}